\documentclass[a4paper,10pt]{article}

\usepackage[a4paper,top=1.9cm,bottom=2.1cm,left=1.7cm,right=1.7cm]{geometry}
\usepackage{graphicx}
\usepackage{float}
\usepackage{subcaption}
\usepackage{amsmath}
\usepackage{amssymb}
\usepackage{amsfonts}
\usepackage{bm}
\usepackage[english]{babel}
\usepackage{csquotes}
\usepackage{authblk}

\usepackage[normalem]{ulem}
\usepackage{xcolor}
\usepackage{microtype}
\usepackage{hyperref}

\allowdisplaybreaks[2]
\definecolor{revisionorange}{RGB}{230,120,0}
\definecolor{revisionblue}{RGB}{0,80,180}
\hypersetup{colorlinks=true,linkcolor=black,citecolor=black,urlcolor=black}

\title{Quartic Scalar Clouds on Fixed Kerr Backgrounds}

\author[1]{Hendrik Mennenga\thanks{\href{mailto:hendrik.mennenga@uni-oldenburg.de}{hendrik.mennenga@uni-oldenburg.de}}}

\affil[1]{Institut f\"ur  Physik, Universit\"at Oldenburg, Postfach 2503,
D-26111 Oldenburg, Germany}

\date{\today}

\begin{document}

\twocolumn[
\begin{@twocolumnfalse}
\maketitle

\begin{abstract}
We study stationary nonlinear clouds of a complex scalar field on fixed Kerr
and Schwarzschild backgrounds, with an unbounded scalar potential. On Kerr backgrounds we find finite-amplitude
Q-clouds satisfying the synchronisation condition \(\omega=m\Omega_H\). They
occupy two-dimensional regions of the \((\Omega_H,M_K)\) plane and reduce to
the corresponding linear clouds as the scalar field amplitude tends to zero. The quartic domain can continue to
\(\Omega_H=0\), where both the synchronised frequency and the Noether charge
vanish and the solutions become static nonlinear scalar clouds on a fixed Schwarzschild background. We also compute the existence domain for a normalised, axion-inspired cosine potential. 
\end{abstract}
\vspace{0.8em}
\end{@twocolumnfalse}
]

\section{Introduction}

The interaction of scalar fields with black holes provides a useful way in which to test the limits of the classical no-hair theorems and to explore possible macroscopic effects of beyonds standard model fields. In asymptotically flat general relativity, standard no-scalar-hair theorems strongly constrain static, regular scalar configurations around black holes \cite{Bekenstein:1995un,Herdeiro:2015waa}. These restrictions can, however, be evaded when some of the assumptions entering such no-hair arguments are not satisfied. A particularly important mechanism is provided by time-dependent complex scalar fields whose stress-energy tensor remains stationary. Examples of static scalar solitons and black holes with scalar hair supported by non-positive scalar potentials, including axially symmetric configurations with an azimuthal winding number, were constructed in Ref.~\cite{Kleihaus:2013tba}. In rotating black hole spacetimes this is closely tied to superradiance. At the threshold condition $\omega=m\Omega_H$, the scalar field is synchronised with the horizon and no net scalar flux crosses it.  Linear massive scalar clouds around Kerr black holes arise at this threshold and form
one-dimensional existence lines in the Kerr parameter space \cite{Hod:2012px,Benone:2014ssa}. 
Their nonlinear continuation allows the construction of Kerr black holes with synchronised scalar hair \cite{Herdeiro:2014goa,Herdeiro:2015gia}. In flat spacetime, complex scalar theories with a global \(U(1)\) symmetry can support non-topological solitons, or Q-balls, whose stability is associated with a conserved Noether charge \cite{Coleman:1985ki,Lee:1991ax}. The standard polynomial Q-ball potential contains a mass term, an attractive quartic interaction, and a stabilising sextic term,

\begin{equation}
U(|\Phi|)
=
\mu^2 |\Phi|^2
-
\lambda |\Phi|^4
+
\beta |\Phi|^6 .
\label{eq:scalar_potential}
\end{equation}

The numerical values of the couplings are model dependent. When such self-interacting scalar fields are placed on a rotating black hole background, the corresponding bound states are known as Q-clouds. 
It was shown that Kerr black holes can support nonlinear Q-clouds which, unlike linear clouds, exist in a two-dimensional region of the Kerr parameter space bounded by the linear existence line and by a minimal horizon angular velocity \cite{Herdeiro:2014pka}. These solutions demonstrate that nonlinear scalar clouds need not be merely zero modes of the linear superradiant instability.
In this work we study a closely related effective model. We consider a complex
scalar field on fixed Kerr and Schwarzschild black-hole backgrounds and focus
on the quartic special case \((\beta=0)\) of Eq.~\eqref{eq:scalar_potential}.
The same structure, namely a positive mass term together with a negative
quartic coupling, appeared already in a discussion of phase transitions
near black-hole horizons ~\cite{Gubser:2005ih}. Gubser
considered uncharged hairy black holes sourced by a real scalar potential
\(V(\phi)=m^2\phi^2/2+\lambda\phi^4/4\), with \(m^2>0\) and \(\lambda<0\).
The later construction of scalarons and static hairy black holes in
Ref.~\cite{Kleihaus:2013tba} used the corresponding complex-field potential
\(U=\mu^2|\Phi|^2-\lambda|\Phi|^4\), which is  the quartic limit
examined here on fixed backgrounds. Gubser emphasised that asymptotically flat uncharged black holes may
support scalar hair when the scalar potential becomes negative at large field
amplitude, and related such unbounded directions to effective potentials in
Calabi--Yau compactifications that can run to negative infinity in suitable
regions of parameter space \cite{Gubser:2005ih,Hertog:2003xg}. The quartic truncation provides a controlled minimal
laboratory for isolating the leading attractive interaction and for studying
how nonlinear clouds behave when the positivity assumptions behind standard
no-hair arguments are relaxed. A complementary small-field motivation comes from axion-like potentials. Their
periodic completion is bounded, but their leading correction to the massive
term is attractive. We therefore also compare the quartic model with the full
normalised, axion-inspired cosine potential. The purpose of the present paper is twofold. First, we isolate the effect of
the leading attractive quartic self-interaction by comparing the quartic
truncation with the previously studied Q-clouds \cite{Herdeiro:2014pka}. Second,
we analyse how the resulting configurations are modified when embedded in
fixed Kerr and Schwarzschild backgrounds. Throughout this work the spacetime geometry is kept fixed, i.e. we neglect the backreaction of the scalar field on the black hole metric. 
\section{Theoretical Framework}
\subsection{The Model}

Throughout this work we use the mostly plus convention and Einstein summation over repeated upper and lower indices. We employ geometrized natural units, \(G=c=\hbar=1\). All dimensionful results are reported through the combinations \(E\mu\), \(J\mu^2\), \(A_{\rm H}\mu^2\), and \(\Omega_{\rm H}/\mu\) with respect to the scalarfield mass $\mu$.

We consider a complex scalar field \(\Phi\) minimally coupled to gravity and invariant under a global \(U(1)\) transformation,
\begin{equation}
\Phi \rightarrow e^{i\alpha}\Phi .
\end{equation}

The scalar-field action is
\begin{equation}
S =
\int d^4x \sqrt{-g}
\left[
- g^{\mu\nu}\partial_\mu \Phi^* \partial_\nu \Phi
- U(|\Phi|)
\right],
\label{action}
\end{equation}
The effective interaction used below keeps the positive mass term and the
attractive quartic self-interaction while omitting higher-order stabilising
terms. Up to normalisation conventions and the replacement of a real scalar
by a complex \(U(1)\)-symmetric field, this is the same quartic sign structure
as in Ref.~\cite{Gubser:2005ih}. The attractive quartic interaction also admits
a complementary small-field motivation from axion-like periodic potentials.
This correspondence applies only near the vacuum. The two theories
remain distinct in their field content, symmetries, and large-field behaviour.
For a canonically normalised real axion-like field \(a\), one may write
\cite{Marsh:2015xka}

\begin{equation}
V(a)
=
\mu^2 f_a^2
\left[
1-\cos\left(\frac{a}{f_a}\right)
\right],
\end{equation}
where \(f_a\) is the axion decay constant. Expanding around \(a=0\) gives
\begin{equation} 
V(a)
=
\frac{1}{2}\mu^2 a^2
-
\frac{1}{24}\frac{\mu^2}{f_a^2}a^4+
\mathcal{O}(a^6).
\end{equation}
The leading nonlinear correction to the quadratic mass term is therefore
attractive. For the parameter choice used below, the corresponding dimensionless
periodic completion is modelled by identifying \(a=\sqrt{2}\,|\Phi|\) and
\(f_a=\mu/\sqrt{12}\). In terms of
\(\hat\phi=\phi/\mu\), this gives
\begin{equation}
\frac{V_{\rm cos}}{\mu^4}
=
\frac{1}{12}
\left[
1-\cos\left(\sqrt{24}\,\hat\phi\right)
\right]
=
\hat\phi^2
-2\hat\phi^4
+\frac{8}{5}\hat\phi^6
+\mathcal{O}(\hat\phi^8).
\label{eq:cosine_completion}
\end{equation}
Since the derivative of the potential
enters the Klein--Gordon equation, we estimate the truncation error from the
relative difference of the scalar force,
\begin{equation}
\epsilon_F(\hat\phi)
=
\left|
\frac{
\frac{d}{d\hat\phi}
\left[
\frac{1}{12}
\left(1-\cos(\sqrt{24}\,\hat\phi)\right)
\right]
-
\frac{d}{d\hat\phi}
\left[
\hat\phi^2-2\hat\phi^4
\right]
}
{
\frac{d}{d\hat\phi}
\left[
\frac{1}{12}
\left(1-\cos(\sqrt{24}\,\hat\phi)\right)
\right]
}
\right| .
\label{eq:taylor_force_error}
\end{equation}
Solving \(\epsilon_F=0.05\) gives
\begin{equation}
\hat\phi=\frac{\phi}{\mu}\simeq 0.2945 .
\end{equation}

This criterion is local in field space, for an
inhomogeneous cloud we apply it to the maximum scalar amplitude
\(\phi_{\rm max}\).
In four spacetime dimensions,
\(
[\Phi]=M
\),
\(
[\mu]=M
\),
\(
[\lambda]=1
\),
and
\(
[\beta]=M^{-2}
\).
We use dimensionless coordinates that are scaled with respect to $\mu$, defining
\(
\bar x^\alpha=\mu x^\alpha
\). Finally we choose 
\[
\mu=1,\qquad \lambda=2,\qquad \beta=0 ,
\]
The full cosine potential is solved separately. Varying the action in Eq.~(\ref{action}) with respect to \(\Phi^*\) gives the nonlinear
Klein Gordon equation
\begin{equation}
\nabla_\mu\nabla^\mu \Phi
=
\frac{\partial U}{\partial |\Phi|^2}\Phi
=
\left(\mu^2-2\lambda|\Phi|^2+3\beta|\Phi|^4\right)\Phi .
\end{equation}

The global \(U(1)\) symmetry gives rise to the conserved Noether current
\begin{equation}
j^\mu
=
-i\left(
\Phi^* \nabla^\mu \Phi
-
\Phi \nabla^\mu \Phi^*
\right),
\qquad
\nabla_\mu j^\mu = 0 .
\end{equation}
The corresponding Noether charge is
\begin{equation}
Q =
\int_\Sigma d^3x \sqrt{-g}\, j^t ,
\end{equation}
where \(\Sigma\) is a spacelike hypersurface. For ordinary Q-balls, the conserved charge \(Q\) labels the solitonic sector and underlies the usual stability criterion, since the energy is minimised at fixed nonzero charge and compared with the energy of \(Q\) free scalar quanta. The stress-energy tensor of the scalar field is
\begin{equation}
\begin{aligned}
T_{\mu\nu}
={}&
\partial_\mu \Phi^* \partial_\nu \Phi
+
\partial_\nu \Phi^* \partial_\mu \Phi
\\
&{}
-
g_{\mu\nu}
\left[
g^{\alpha\beta}
\partial_\alpha \Phi^* \partial_\beta \Phi
+
U(|\Phi|)
\right],
\end{aligned}
\end{equation}
therefore the mass energy $E$ and angular momentum $J$ of the system are given by 
\begin{equation}
\begin{aligned}
E ={}& -2\pi \int_{r_H}^{\infty} dr
\int_{0}^{\pi} d\theta \, \sqrt{-g} \, T^{t}_{t},
\end{aligned}
\end{equation}

\begin{equation}
\begin{aligned}
J ={}& 2\pi \int_{r_H}^{\infty} dr
\int_{0}^{\pi} d\theta \, \sqrt{-g} \, T^{t}_{\varphi}.
\end{aligned}
\end{equation}
For the scalar field we use the harmonic ansatz

\begin{equation}
\Phi(t,r,\theta,\varphi)
=
\phi(r,\theta)
e^{i(m\varphi-\omega t)} ,
\end{equation}
where \(m\in\mathbb{Z}\) is the azimuthal winding number and \(\omega\) is the scalar field frequency. The scalar amplitude \(\phi(r,\theta)\) is taken to be real. With this ansatz the scalar field is explicitly time- and azimuth-dependent, but all physical observables constructed from \(|\Phi|\), \(j^\mu\) and \(T_{\mu\nu}\) are stationary and axisymmetric, because they only depend on \(|\Phi|^2\) and so the phase cancels. In the limit of a linear cloud the Klein Gordon equation on the Kerr background
is separable (in Boyer--Lindquist coordinates) \cite{Dolan:2007mj}. With the mode decomposition
\begin{equation}  
\Phi
=
e^{-i\omega t}e^{im\varphi}
S_{\ell m}(\theta)R_{n\ell m}(r),
\end{equation}
the angular dependence is described by spheroidal harmonics, while the radial
equation becomes an eigenvalue problem for bound states. The resulting linear
clouds are therefore labelled by \enquote{quantum numbers}  \((n,\ell,m)\), where
\(n\) counts the radial nodes, \(\ell\) is the angular harmonic index, and
\(m\) is the azimuthal harmonic index. Imposing the synchronisation condition
\(\omega=m\Omega_H\) selects one-dimensional existence lines in the
two-dimensional Kerr parameter space for each fixed set \((n,\ell,m)\)
\cite{Hod:2012px,Hod:2013zza,Benone:2014ssa,Herdeiro:2014pka}. The asymptotic form of the scalar equation is independent of the nonlinear
interaction,
\begin{equation}
    \left(\nabla^2-\kappa^2\right)\phi\simeq 0,
    \qquad
    \kappa^2=\mu^2-\omega^2 .
    \label{eq:flat_localisation_scale}
\end{equation}
Exponential localisation therefore requires \(\omega<\mu\). In the units used
for our solutions, \(\mu=1\), so the upper endpoint is $\omega^{\rm(max)}=1$.

\subsection{The Ansatz}

For computation we use quasi isotropic coordinates \cite{Kleihaus:2000kg} because they place the horizon at a regular
inner boundary of the numerical domain, where the synchronised scalar field can
be imposed by a simple Neumann condition. 
\begin{equation}
\begin{aligned}
ds^2={}&
-f(r,\theta)\,dt^2
+
\frac{p(r,\theta)}{f(r,\theta)}
\left(dr^2+r^2d\theta^2\right)
\\
&+
\frac{l(r,\theta)}{f(r,\theta)}
r^2\sin^2\theta
\left(
d\varphi-\frac{\omega_K(r,\theta)}{r}\,dt
\right)^2 .
\end{aligned}
\label{eq:metric_ansatz}
\end{equation}

The metric functions \(f,p,l\), and \(\omega_K\) depend only on \(r\) and \(\theta\) and can be written as 
\begin{equation}
\begin{aligned}
f={}&\left(1-\frac{r_{\rm H}^2}{r^2}\right)^2\frac{F_1}{F_2},
\\
l={}&\left(1-\frac{r_{\rm H}^2}{r^2}\right)^2,
\\
p={}&\left(1-\frac{r_{\rm H}^2}{r^2}\right)^2\frac{F_1^2}{F_2},
\\
\omega_K={}&\frac{2M_K\sqrt{M_K^2-4r_{\rm H}^2}}{r^2}
\frac{(1+\frac{M_K}{r}+\frac{r_{\rm H}^2}{r^2})}{F_2},
\end{aligned}
\label{Kerr}
\end{equation}
 where
\begin{align}
F_1={}&
\frac{2M_K^2}{r^2}
+\left(1-\frac{r_{\rm H}^2}{r^2}\right)^2
+\frac{2M_K}{r}\left(1+\frac{r_{\rm H}^2}{r^2}\right)
\nonumber\\
&-\frac{M_K^2-4r_{\rm H}^2}{r^2}\sin^2\theta,
\label{functions-Kerr1}
\\
F_2={}&
\left(\frac{2M_K^2}{r^2}
+\left(1-\frac{r_{\rm H}^2}{r^2}\right)^2
+\frac{2M_K}{r}\left(1+\frac{r_{\rm H}^2}{r^2}\right)\right)^2
\nonumber\\
&-\left(1-\frac{r_{\rm H}^2}{r^2}\right)^2
\frac{M_K^2-4r_{\rm H}^2}{r^2}\sin^2\theta.
\label{functions-Kerr2}
\end{align}
In quasi isotropic coordinates, the Kerr background exhibits a two branch
structure when the horizon angular velocity \(\Omega_H\) is kept fixed and the
quasi-isotropic horizon radius \(r_H\) is varied. This structure follows from
the relation
\begin{equation}
\Omega_H
=
\frac{\sqrt{M_K^2-4r_H^2}}
     {2M_K(M_K+2r_H)} ,
\end{equation}
which does not determine the Kerr mass \(M_K\) uniquely. Introducing the dimensionless quantities
\begin{equation}
x=\frac{r_H}{M_K},
\qquad
U=\Omega_H r_H,
\end{equation}
gives the cubic equation
\begin{equation}
x^3-\frac{1}{2}x^2+4U^2x+2U^2=0 .
\end{equation}
For \(0<r_H<r_H^{\rm max}\), this equation possesses two physically
admissible positive roots and hence two different Kerr masses,
\(M_K^{\rm low}\) and \(M_K^{\rm up}\), for the same boundary data
\((r_H,\Omega_H)\). The first branch starts at \(r_H=0\) with \(M_K^{\rm low}\rightarrow0\), and
therefore approaches Minkowski spacetime. As \(r_H\) increases, this branch
reaches a maximal horizon radius \(r_H^{\rm max}\), where the two positive
roots merge. The second branch extends back from this turning point toward
\(r_H=0\). Along this branch, the Kerr mass approaches
\begin{equation}
M_K^{\rm up}\longrightarrow\frac{1}{2\Omega_H},
\end{equation}
and the extremal Kerr limit is reached. Thus, \(r_H\rightarrow0\) has two
different interpretations in quasi-isotropic coordinates. It describes
Minkowski spacetime on the first branch and an extremal Kerr black hole on the
second branch. This two-branch structure is a consequence of parametrising
the Kerr family by \((r_H,\Omega_H)\).
The partial differential equation solved numerically is obtained by inserting
the ansatz into the Klein--Gordon equation on the quasi-isotropic Kerr
background. We define
\[
    U_H(r)=1-\frac{r_H^2}{r^2},
    \qquad
    W(r,\theta)=\frac{\omega_K(r,\theta)}{r},
\]
where \(\omega_K\) is the metric function in Eq.~\eqref{eq:metric_ansatz}.
With \(F_1\) and \(F_2\) denoting the Kerr background functions, the reduced
scalar equation in the physical coordinates (\(r,\theta\)) can be written as
\begin{equation}
\begin{aligned}
0={}&
r^2\sin^2\theta\,\partial_r^2\phi
+\frac{r\sin^2\theta}{U_H}
\left(2U_H+r\partial_rU_H\right)\partial_r\phi
\\
&+
\sin^2\theta\,\partial_\theta^2\phi
+\sin\theta\cos\theta\,\partial_\theta\phi
-\frac{m^2F_1^2}{F_2}\,\phi
\\
&+
r^2\sin^2\theta
\left[
-F_1\frac{\partial U}{\partial |\Phi|^2}
+\frac{F_2}{U_H^2}\left(\omega-mW\right)^2
\right]\phi .
\end{aligned}
\label{eq:reduced_scalar_equation}
\end{equation}
In the flat-space limit one sets \(r_H=0\), \(F_1=F_2=1\), \(U_H=1\), and
\(W=0\). The same equation then reduces to the spinning Q-ball equation in
Minkowski spacetime. For the spinning solutions
considered here, \(m\neq0\), regularity requires
\(\phi|_{r=0}=0\) and \(\phi|_{\theta=0}=0\). Localisation imposes
\(\phi|_{r\rightarrow\infty}=0\), and equatorial reflection symmetry gives
\(\partial_\theta\phi|_{\theta=\pi/2}=0\). For \(m=0\), the regularity
conditions at the origin and on the axis are replaced by the corresponding
Neumann conditions. On a black-hole background the origin is replaced by the
horizon. The curved-space boundary conditions are
\(\partial_r\phi|_{r=r_H}=0\), \(\phi|_{r\rightarrow\infty}=0\),
\(\phi|_{\theta=0}=0\) for \(m\neq0\), and
\(\partial_\theta\phi|_{\theta=\pi/2}=0\), together with the synchronisation
condition in Eq.~\eqref{eq:synchronisation_condition}. The  energy- and angular momentum density take the form

\begin{equation}
\begin{aligned}
T^t{}_{\varphi}
&=
\frac{2mF_2\phi^2(\omega-mW)}
     {U_H^2F_1},
\\
-T^t{}_{t}
&=
\frac{m^2F_1\phi^2}{r^2F_2\sin^2\theta}
+U(\phi)
\\
&\quad
+\frac{r^2(\partial_r\phi)^2+(\partial_\theta\phi)^2}{r^2F_1}
+\frac{F_2\phi^2(\omega^2-m^2W^2)}{U_H^2F_1}.
\end{aligned}
\label{eq:reduced_stress_tensor}
\end{equation}
The quantities in Eq.~\eqref{eq:reduced_stress_tensor} are associated with
the Killing vectors \(\partial_t\) and \(\partial_\varphi\), respectively.
\(-T^t{}_{t}\) is the density associated with the conserved
Killing energy. On a rotating background it is not, however, the energy
density measured locally by a physical observer. A natural local observer is a zero-angular-momentum observer (ZAMO). ZAMOs
have vanishing angular momentum, \(n_\varphi=0\), but are dragged by the
geometry and therefore co-rotate with the local angular velocity \(W\).
Their unit four-velocity is
\begin{equation}
    n^\mu
    =
    \frac{\sqrt{F_2}}{U_H\sqrt{F_1}}
    \left(\partial_t+W\partial_\varphi\right)^\mu .
    \label{eq:zamo_four_velocity}
\end{equation}
The energy density measured by a ZAMO is
\begin{equation}
\begin{aligned}
\rho_{\rm ZAMO}
&=
T_{\mu\nu}n^\mu n^\nu
=
\frac{m^2F_1\phi^2}{r^2F_2\sin^2\theta}
+U(\phi)
\\
&\quad
+\frac{r^2(\partial_r\phi)^2+(\partial_\theta\phi)^2}{r^2F_1}
+\frac{F_2\phi^2(\omega-mW)^2}{U_H^2F_1}.
\end{aligned}
\label{eq:zamo_energy_density}
\end{equation}
Thus, the locally measured energy density differs from the Killing energy
density by the frame dragging contribution,
\begin{equation}
    \rho_{\rm ZAMO}
    =
    -T^t{}_{t}
    -W T^t{}_{\varphi}.
    \label{eq:zamo_killing_energy_relation}
\end{equation}
Regularity at the event horizon requires the scalar phase to be synchronised with the horizon generator $\chi^\mu\partial_\mu\Phi=0$. For a rotating black hole with angular velocity \(\Omega_H\), this gives
\begin{equation}
\omega = m\Omega_H .
\label{eq:synchronisation_condition}
\end{equation}
This condition ensures that there is no net scalar flux through the horizon. In the Kerr case it is the familiar superradiant threshold condition. In the Schwarzschild limit, where \(\Omega_H=0\), it instead imposes
\begin{equation}
\omega=0 .
\label{eq:schwarzschild_zero_frequency}
\end{equation}
This limit is normally incompatible with the frequency window of standard Q-balls based on a stabilised sextic potential. In the present model, however, the absence of a positive lower frequency bound allows one to construct scalar clouds around Schwarzschild black holes.

\subsection{Numerical Method}

The field equation(s) are solved as a boundary value problem for a system of
nonlinear, elliptic partial differential equations in the two variables
\((r,\theta)\). To treat the semi-infinite radial domain,
\(
r \in [r_H,\infty),
\)
we introduce a compactified coordinate
\(
x \in [0,1]
\)
via the mapping
\begin{equation}
r(x) = \frac{r_H + c\, x}{1 - x},
\end{equation}
where \( r_H \) denotes the horizon radius and \( c \) is a positive
compactification parameter controlling the radial resolution.
This transformation maps spatial infinity to the finite boundary \(x=1\),
while the horizon is located at \(x=0\).
The angular coordinate is taken as
\(
\theta \in [0,\pi/2],
\)
so that the full domain is reconstructed by reflection symmetry.
With this choice, the computational domain becomes the rectangle
\((x,\theta)\in[0,1]\times[0,\pi/2]\). The derivatives are represented by sixth order finite difference matrices on a
uniform \(N_x\times N_\theta\) grid with \(N_x=300\) and \(N_\theta=80\). The
 system is solved either by a damped Newton method with an
explicit sparse Jacobian or, when the Newton step is poorly conditioned, by a
least squares variant using the same residual.  Families of solutions are obtained by continuation in \(\Omega_H\) or \(r_H\), using the preceding converged solution as the initial guess. Typical residuals are of order \(10^{-6}\) for the
background solutions and \(10^{-8}\) for the flat-space
solutions. As an independent check we use a virial identity obtained from the
horizon-adapted scaling
\begin{equation}  
\phi_\lambda(r,\theta)=
\phi(r_H+\lambda(r-r_H),\theta).
\end{equation}
The relative violation is of order \(10^{-5}\) or smaller.

\section{Flat-Space Solutions}

\begin{figure}[H]
    \centering
    \includegraphics[width=\columnwidth]{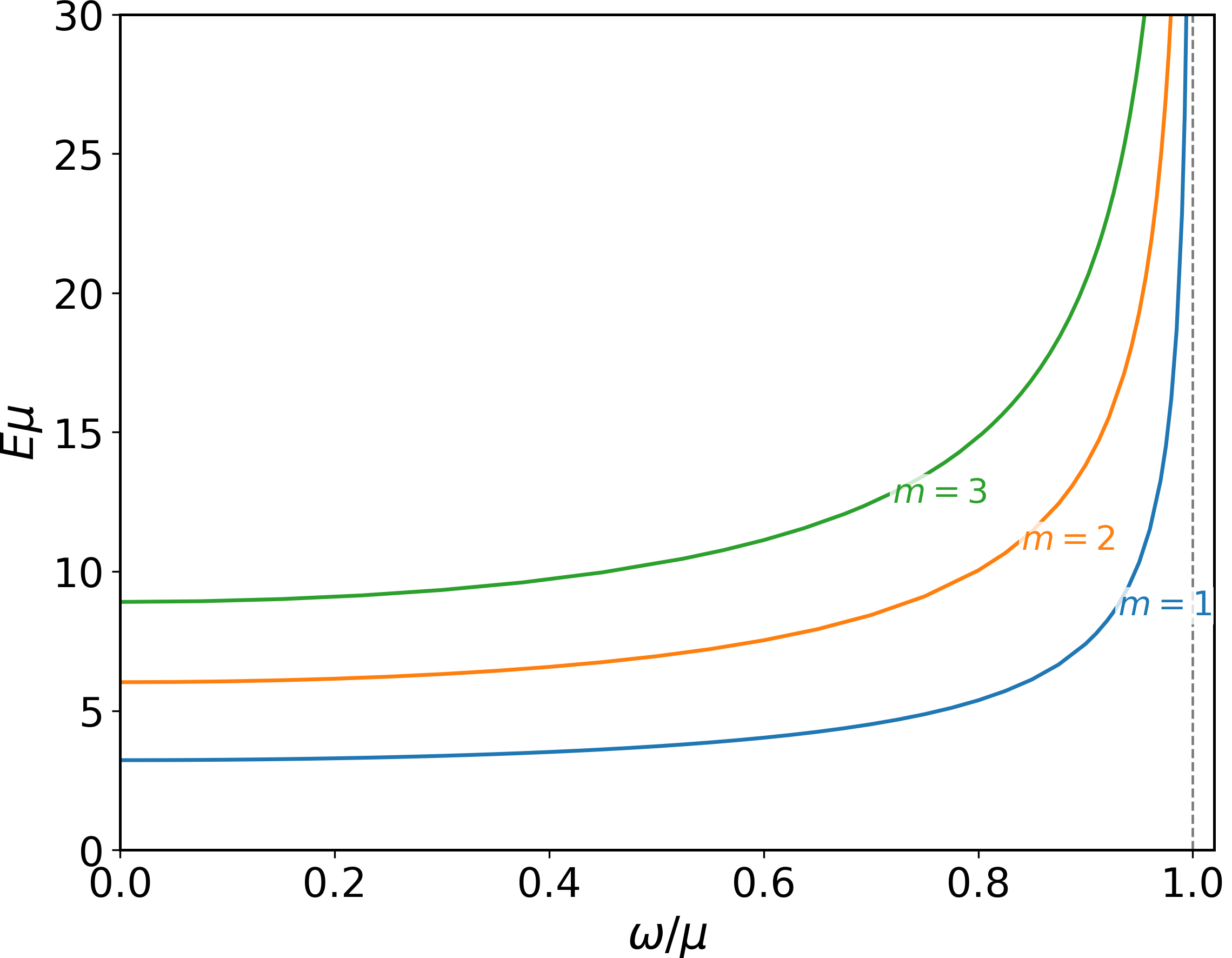}
    \caption{Dimensionless energy \(E\mu\) of the
    flat space quartic solutions as functions of dimensionless scalarfield frequency $\omega/\mu$  for winding numbers \(m=1,2,3\). The dashed vertical lines denote the localisation threshold.}
    \label{fig:flat_global_charges}
\end{figure}

\begin{figure}[H]
    \centering
    \includegraphics[width=\columnwidth]{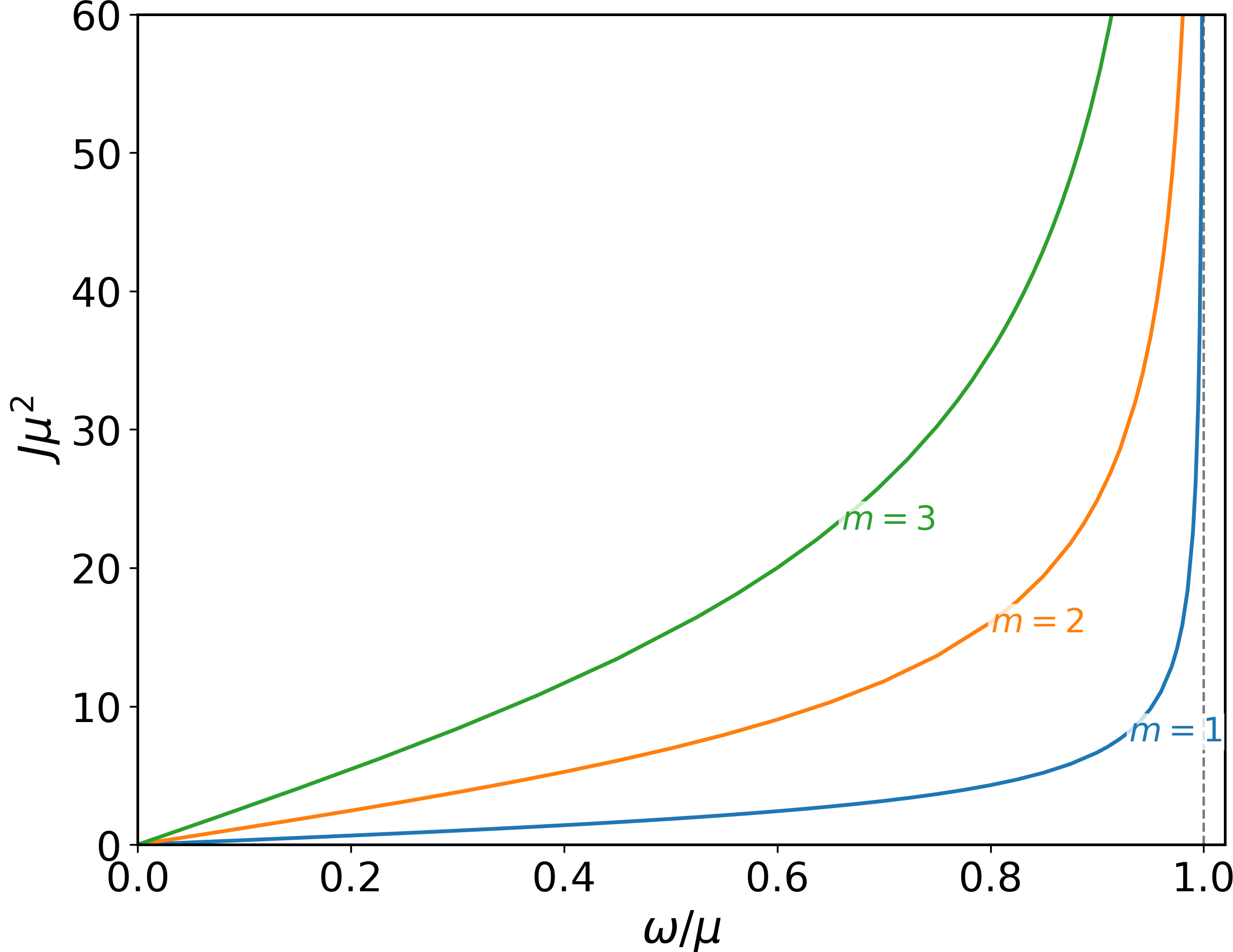}
    \caption{Same as Fig. \ref{fig:flat_global_charges} for dimensionless angular momentum \(J\mu^2\).}
    \label{fig:flat_global_angular_momentum}
\end{figure}

In the limit of Minkowski spacetime, the system reduces to spinning Q-ball-type 
solutions. Spinning Q-balls were first constructed
numerically in \cite{Volkov:2002aj}, and were subsequently
studied as the flat space limit of rotating boson stars \cite{Kleihaus:2005me}. Conversely, \cite{Herdeiro:2014pka} showed
that spinning Q-balls based on a stabilised polynomial potential survive when
a Kerr horizon is inserted at their centre and become synchronised Q-clouds, which need to obey Eq.~\eqref{eq:synchronisation_condition}. This interpretation has since been used more generally
as a probe for black holes with synchronised hair \cite{Herdeiro:2017oyt}.  Figs. \ref{fig:flat_global_charges} and \ref{fig:flat_global_angular_momentum} show the energy and angular momentum of
the quartic flat space solutions for \(m=1,2,3\). The dashed vertical lines
mark the corresponding localisation thresholds. The energy
and angular momentum diverge when these thresholds are approached. This
behaviour is the flat-space counterpart of the large-charge regime familiar
from Q-ball solutions \cite{Coleman:1985ki,Lee:1991ax}, although the present quartic model should not be interpreted as a stable Coleman Q-ball because its potential is unbounded from below.

\begin{figure}[H]
    \centering
    \includegraphics[width=\columnwidth]{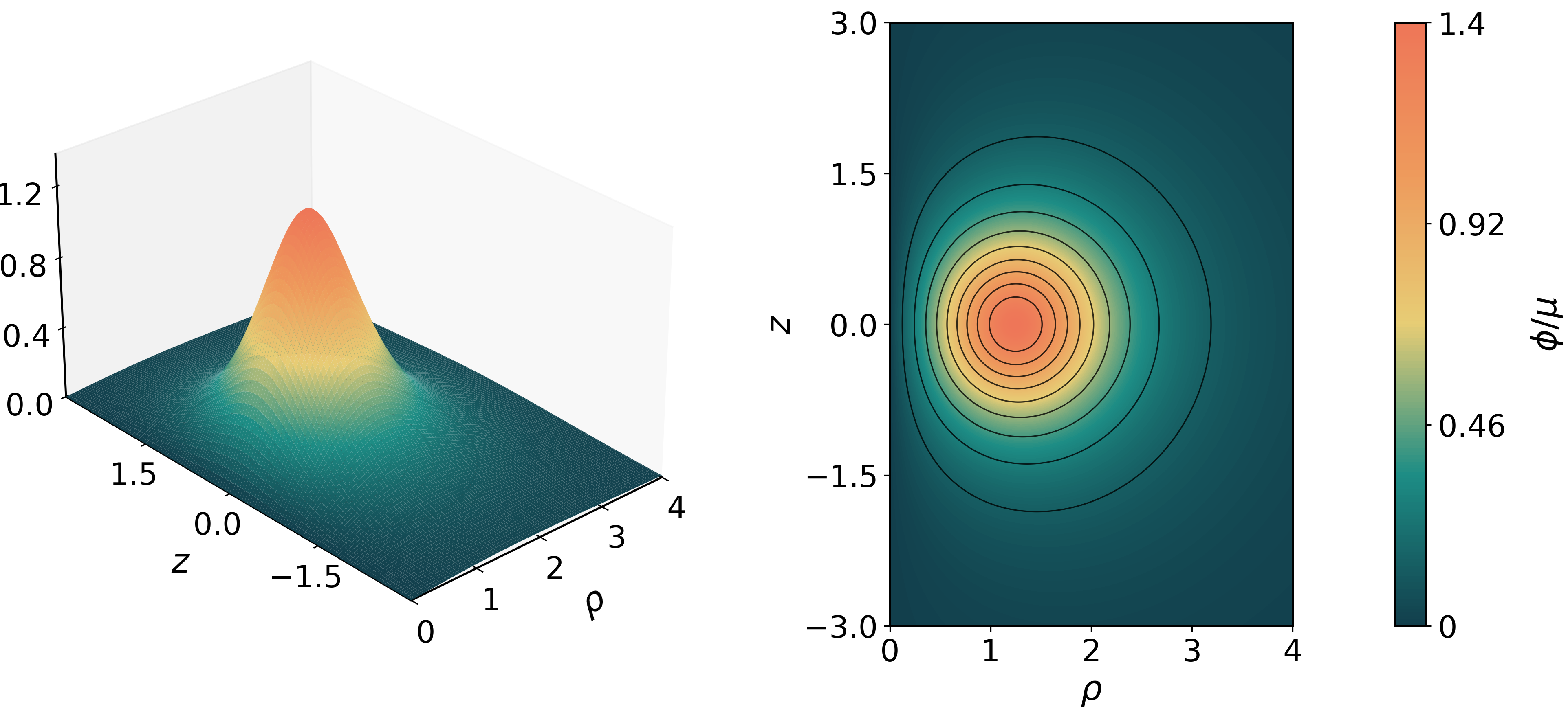}
    \caption{Scalar amplitude $\phi / \mu$ of the flat space quartic solution with
    \(m=1\) and \(\Omega_H=0.5\). The left panel shows a three dimensional
    representation of the meridional profile and the right panel its contour
    plot in cylindrical coordinates.}
    \label{fig:flat_phi_profile}
\end{figure}

\begin{figure}[H]
    \centering
    \includegraphics[width=\columnwidth]{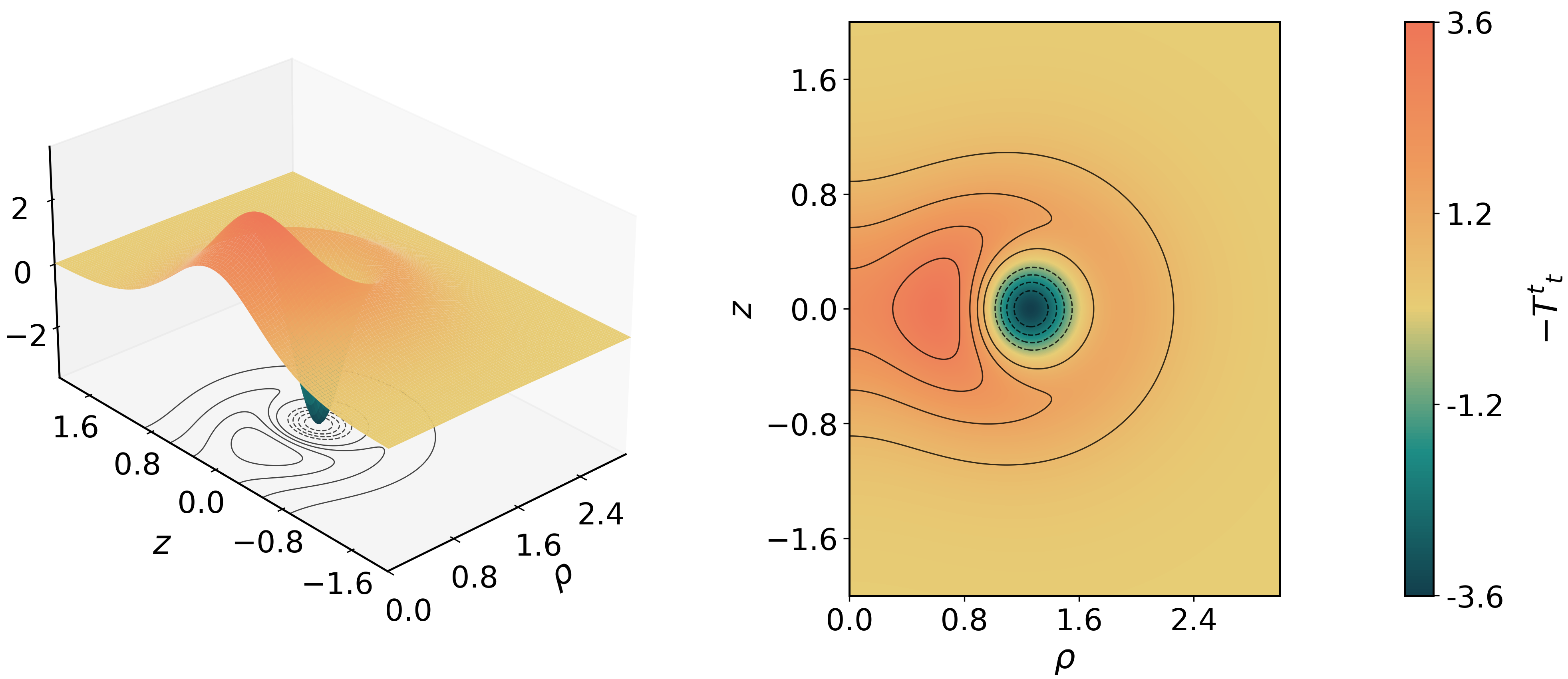}
    \caption{Same as Fig.~\ref{fig:flat_phi_profile} for the Killing energy density \(-T^t{}_{t}\).}
    \label{fig:flat_energy_density}
\end{figure}

\begin{figure}[H]
    \centering
    \includegraphics[width=\columnwidth]{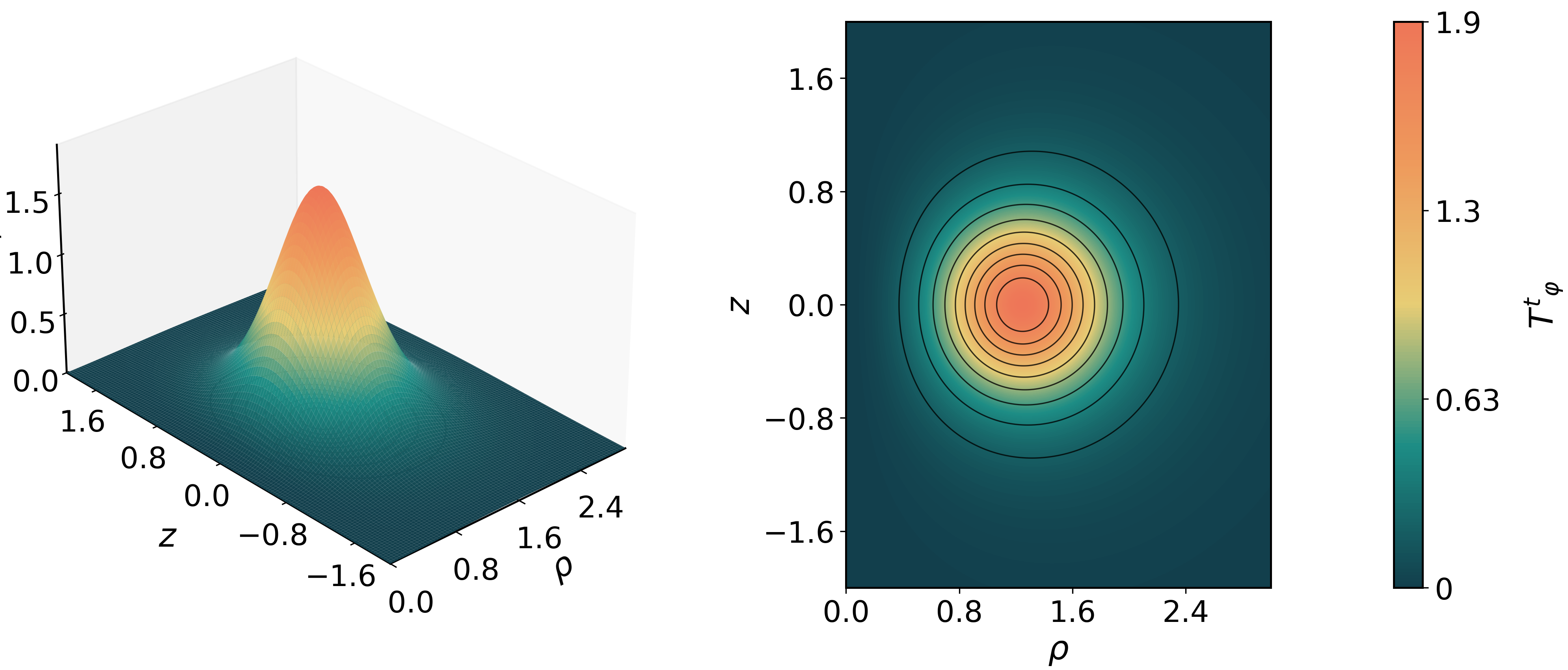}
    \caption{Same as Fig.~\ref{fig:flat_phi_profile} for the angular momentum density \(T^t{}_{\varphi}\).}
    \label{fig:flat_angular_momentum_density}
\end{figure}

For the
quartic truncation, the ratio relevant for the standard Q-ball
existence argument is
\begin{equation}
    \frac{U(\phi)}{\phi^2}=\mu^2-\lambda\phi^2 .
\end{equation}
It is unbounded from below, and hence the quartic model does not impose a
positive lower bound on \(\omega\). Our numerical branches consequently extend
to \(\omega=0\), while the scalar profile and its total energy remain nonzero. The
Noether charge and angular momentum vanish.
Their localisation is supported by the
unbounded attractive quartic interaction rather than by energy minimisation at
fixed nonzero charge. For comparison, consider the standard stabilised case,
Eq.~\eqref{eq:scalar_potential} with \(\beta>0\).
The sextic term makes the potential bounded from below, as required for an
ordinary stable Q-ball model \cite{Heeck:2022iky}. In this case the frequency
window is
\begin{equation}
    \omega_{\rm min}^2
    =
    \min_{\phi\neq 0}\frac{U(\phi)}{\phi^2}
    =
    \mu^2-\frac{\lambda^2}{4\beta},
    \qquad
    \omega_{\rm max}^2=\mu^2 .
    \label{eq:sextic_frequency_window}
\end{equation}
For a non degenerate stable potential one has \(\omega_{\rm min}>0\), and the
zero frequency endpoint is excluded. The tuned sextic choice
\(U(\phi)=\phi^2(1-\phi^2)^2\) is exceptional. In this case the formal minimum
of \(U(\phi)/\phi^2\) is zero, because the potential has a degenerate vacuum at
nonzero field amplitude. However, \(\omega=0\) is not reached as a regular
finite charge Q-ball, it appears only as a singular
thin wall limiting point. The quartic truncation is different. It has no lower
frequency bound because the potential is unbounded from below. This is  why it admits the
finite energy \(\omega=0\) configurations found here.
The local structure of a generic solution is shown in
Figs.~\ref{fig:flat_phi_profile}--\ref{fig:flat_angular_momentum_density}. We
choose \(m=1\) and \(\Omega_H=0.5\), and use cylindrical coordinates
\((\rho,z)\). The scalar amplitude vanishes on the symmetry axis, at the
origin and at spatial infinity. Its maximum is located away from the axis, so
that a rotation of the meridional profile around the \(z\)-axis produces toroidal distribution of an even-parity spinning Q-ball. This
agrees with the established morphology of spinning Q-balls
\cite{Volkov:2002aj,Kleihaus:2005me}. The energy density in Fig.~\ref{fig:flat_energy_density} is not
everywhere positive. In particular, the configuration develops a negative
core surrounded by a positive contribution. This is a consequence of the attractive quartic term. The integrated energy remains
positive, as shown in Fig.~\ref{fig:flat_global_charges}. Finally,
Fig.~\ref{fig:flat_angular_momentum_density} shows that the angular momentum
density is localised in the same toroidal region as the scalar field. As for
ordinary spinning Q-balls, the global quantities obey the quantisation
relation
\begin{equation}
    J=mQ .
    \label{eq:angular_momentum_quantisation}
\end{equation}

\section[Static scalar clouds OmegaH=0]{Static scalar clouds $\Omega_H=0$}

\begin{figure}[H]
\centering
\includegraphics[width=\columnwidth]{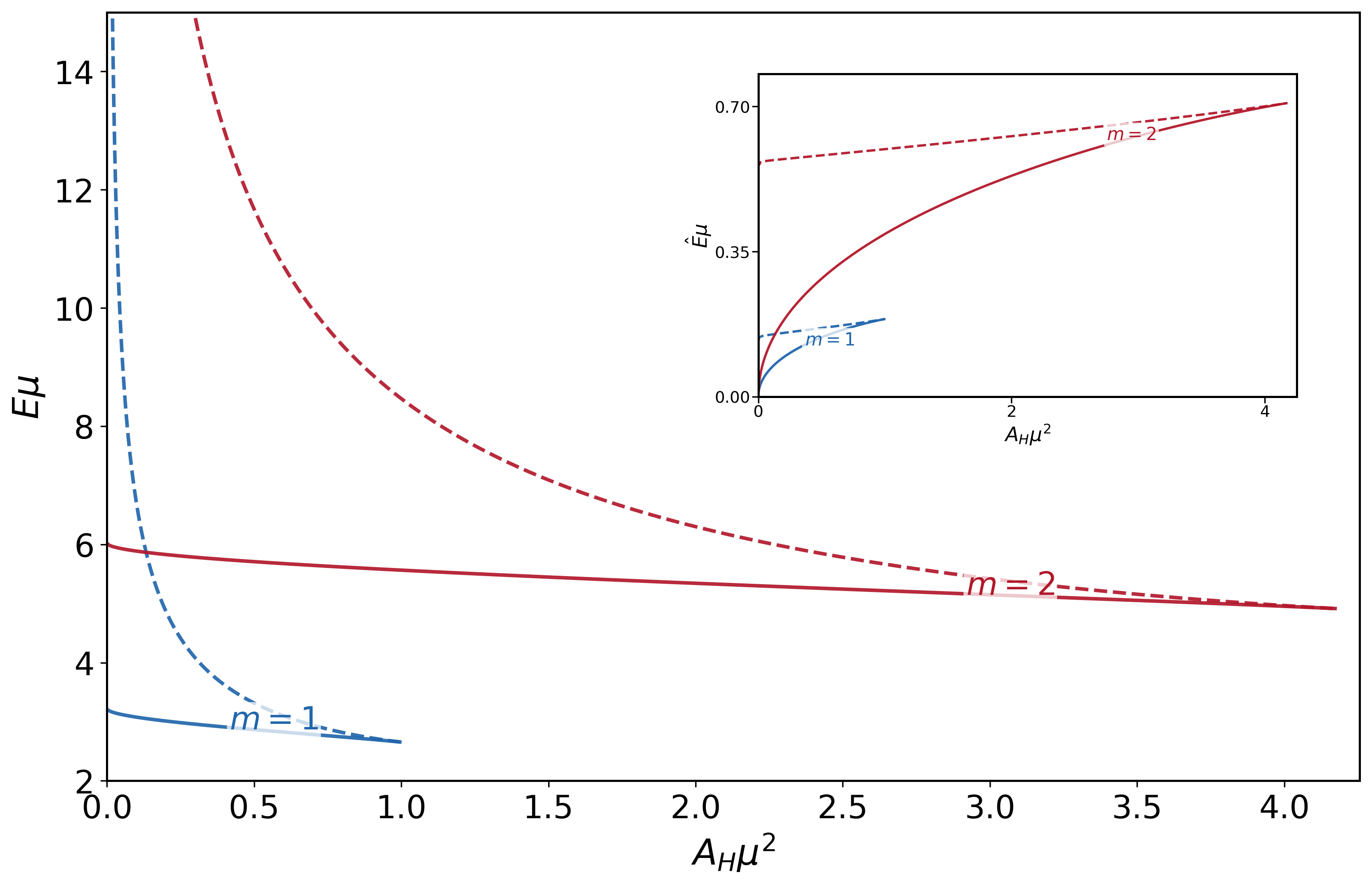}
\caption{Dimensionless energy $E\mu$ as a function of the dimensionless
horizon area $A_H\mu^2$ for $\Omega_H=0$ and winding numbers $m=1,2$.
Solid and dashed curves denote the two branches. The inset shows the
rescaled energy $\hat{E}\mu=(E\mu)r_H$, where $r_H$ is the dimensionless
horizon radius.}
\label{fig:static_cloud_energy}
\end{figure}

Setting the horizon angular velocity to zero reduces the Kerr background to
a Schwarzschild black hole. The synchronisation condition
Eq.~\eqref{eq:synchronisation_condition} then enforces
Eq.~\eqref{eq:schwarzschild_zero_frequency}. In quasi-isotropic coordinates,
the metric follows from Eq.~\eqref{eq:metric_ansatz} by setting
$\omega_K=0$,

\begin{equation}
\begin{aligned}
ds^2={}&-\frac{(1-\frac{r_H}{r})^2}
{(1+\frac{r_H}{r})^2} dt^2
+\left(1+\frac{r_H}{r}\right)^4
\\
&{}\times
\left[dr^2+r^2\left(d\theta^2+\sin^2\theta\,d\varphi^2\right)\right].
\end{aligned}
\label{eq:schwarzschild_quasi_isotropic_metric}
\end{equation}
The scalar field is time independent,
\begin{equation}
\Phi(r,\theta,\varphi)
=
\phi(r,\theta)e^{im\varphi},
\end{equation}
but retains a nontrivial azimuthal phase for $m\neq0$. Its modulus and
stress-energy tensor are static and axisymmetric. Since the Schwarzschild
background has no $t$-$\varphi$ mixing, the temporal component of the
Noether current and the angular momentum density vanish. Consequently,
\begin{equation}
Q=0,
\qquad
J=0 .
\end{equation}
Unlike linear Kerr clouds, these configurations do not rely on synchronised
rotation \cite{Hod:2012px,Herdeiro:2014goa}. Standard Q-clouds supported by a
bounded quartic-plus-sextic potential also exist only above a minimum horizon
angular velocity \cite{Herdeiro:2014pka}. The static configurations found here
are therefore intrinsically nonlinear. This follows directly from the scalar equation. Multiplying the static
Klein--Gordon equation by $\Phi^*$ and integrating over the exterior region
gives
\begin{equation}
\int_{\Sigma} d^3x\sqrt{-g}
|\nabla\Phi|^2
+
\mu^2|\Phi|^2
-
2\lambda|\Phi|^4
=
0
\label{eq:static_virial_balance}
\end{equation}
where the boundary terms vanish for a regular horizon and a localised field.
For a free massive scalar field, all contributions are non-negative, so only
the trivial static solution exists. In the quartic model, the attractive
self-interaction can balance the gradient and mass terms and thereby support
static clouds, i.e. the potential violates the positivity assumptions
entering standard no-scalar-hair arguments
\cite{Bekenstein:1995un,Herdeiro:2015waa}. Fig. \ref{fig:static_cloud_energy} shows the energy as a function of the
horizon area. For $\Omega_H=0$, one has $M_K=2r_H$, and hence
$A_H=16\pi M_K^2=64\pi r_H^2$. For each winding number, two branches meet at
a maximal horizon area. The solid branch approaches the corresponding static
flat-space solution as $A_H\rightarrow0$, while its energy decreases when a
Schwarzschild horizon is inserted. The second branch returns towards
$A_H\rightarrow0$, with diverging energy. The finite limit of
$\hat{E}r_H$ indicates the scaling $E\sim r_H^{-1}$. Increasing $m$ raises
both the energy scale and the maximal horizon area. The same qualitative
behaviour was found for $m=3,4,5$. Related static configurations supported by non-positive scalar potentials
were constructed in Ref.~\cite{Kleihaus:2013tba}. The solutions found here
are closely related to this scalaron mechanism, but arise on a fixed
background as the zero-angular-velocity and zero-frequency endpoints of
synchronised Kerr Q-cloud branches.

\section{Quartic Q-Clouds}

\begin{figure}[H]
    \centering
    \includegraphics[width=\columnwidth]{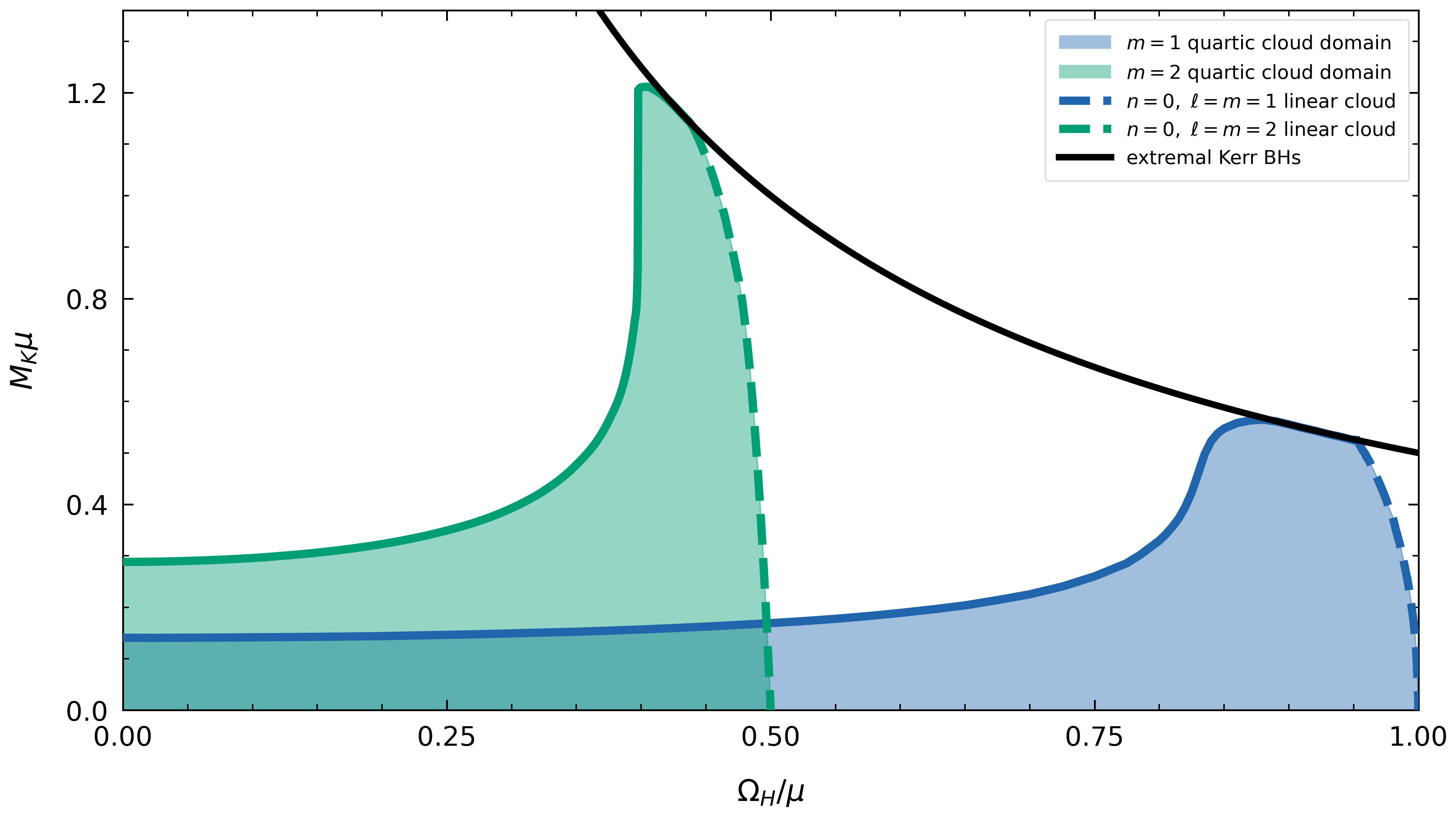}
\caption{Existence domains of quartic scalar clouds in the Kerr parameter
space. The black curve denotes extremal Kerr black holes. The dashed curves
are the fundamental linear-cloud existence lines with radial node number
\(n=0\) and \(\ell=m=1,2\). The blue and green regions indicate the Kerr
backgrounds supporting nonlinear quartic clouds with azimuthal winding
numbers \(m=1\) and \(m=2\), respectively.}
\label{fig:existence_domain}
\end{figure}

The attractive quartic interaction gives rise to
nonlinear Q-clouds on Kerr backgrounds. These configurations
obey the synchronisation condition Eq.~\eqref{eq:synchronisation_condition}
and are therefore stationary no flux solutions. At the boundary of the
nonlinear domain the scalar amplitude tends to zero, and the solutions reduce
to the corresponding linear Kerr clouds. This is the same mechanism used in \cite{Herdeiro:2014pka}. Linear massive scalar
clouds occupy one-dimensional existence lines in the Kerr parameter space
\cite{Hod:2012px,Benone:2014ssa}. For fixed azimuthal winding number \(m\),
the relevant fundamental line has radial node number \(n_r=0\) and
\(\ell=m\). When an attractive self-interaction is included, finite-amplitude
solutions can exist away from this line, so the one-dimensional linear
existence line opens into a two-dimensional nonlinear domain. For comparison, Ref.~\cite{Herdeiro:2014pka} considered
Eq.~\eqref{eq:scalar_potential} with
\(\mu^2=1.1\), \(\lambda=2\), and \(\beta=1\). For this parameter choice
the potential is bounded from below and strictly non-negative, since
\(U(\phi)/\phi^2=(\phi^2-1)^2+0.1\), and the associated flat-space Q-balls
have a finite frequency window. Through
Eq.~\eqref{eq:synchronisation_condition}, its positive lower endpoint becomes
a lower bound on the horizon angular velocity,
\(\Omega_H^{\rm(min)}=\omega_{\rm min}/m\). The sextic Q-cloud domains are
therefore bounded by a minimum \(\Omega_H\), by the linear cloud existence
line and, as it turns out, by the extremal Kerr curve. In contrast,
for our choice the quartic families continue to \(\Omega_H=0\), where they connect to the static
scalar clouds discussed in the previous section. We have compared the two potentials numerically for different
synchronised frequencies and horizon sizes. In both cases a small horizon
excises the central part of the corresponding flat-space configuration. For the sextic
potential, the Q-cloud family connected to a flat-space Q-ball can be
continued smoothly along the two geometric Kerr branches, \(M_K^{\rm low}\) and
\(M_K^{\rm up}\). In the numerical domain explored here, the quartic branches approach the
extremal Kerr curve only in limited regions of parameter space. Away from these
regions the continuation does not reach the maximal horizon area of the Kerr
family. Instead, the numerical branch stops earlier and bends back while still
on the \(M_K^{\rm low}\) branch. At fixed \(\Omega_H\), where the extremal Kerr limit is not reached, the lower
geometric branch starts from the Minkowski limit and extends to a maximal
horizon radius. The continuation then folds onto a second branch, where the
scalar energy grows rapidly and diverges as the input parameter \(r_H\rightarrow0\). The sextic potential yields a nonnegative ZAMO energy density, whereas the quartic configurations considered here contain regions of negative
ZAMO energy density, because the negative potential term can locally dominate the remaining contributions. Fig. \ref{fig:existence_domain} shows the resulting domains in the Kerr
parameter space $(\Omega_H,M_K)$. For each value of \(m\) displayed, there is a finite region of
Kerr backgrounds supporting nonlinear quartic clouds. The dashed curves are the
corresponding fundamental linear-cloud existence lines with \(n=0\) and
\(\ell=m\). Along these boundaries, the scalar amplitude tends to zero and the nonlinear
solution reduces to a linear cloud, i.e. a stationary bound state solution of
the linearised Klein Gordon equation on the fixed black hole background,
existing at the synchronisation threshold. The shaded domains therefore
represent the nonlinear continuation of these one dimensional linear
existence lines. The relation to the extremal Kerr curve is different from the bounded sextic
case. In Ref.~\cite{Herdeiro:2014pka}, the entire Q-Cloud domain is bounded
by the extremal Kerr line. In the quartic model, the nonlinear domains for the
different \(m\) do not generically extend all the way to extremality. They
approach the extremal Kerr boundary only in limited regions of the
parameter space. Thus extremality is not a global boundary.

\begin{figure}[H]
    \centering
    \includegraphics[width=\columnwidth]{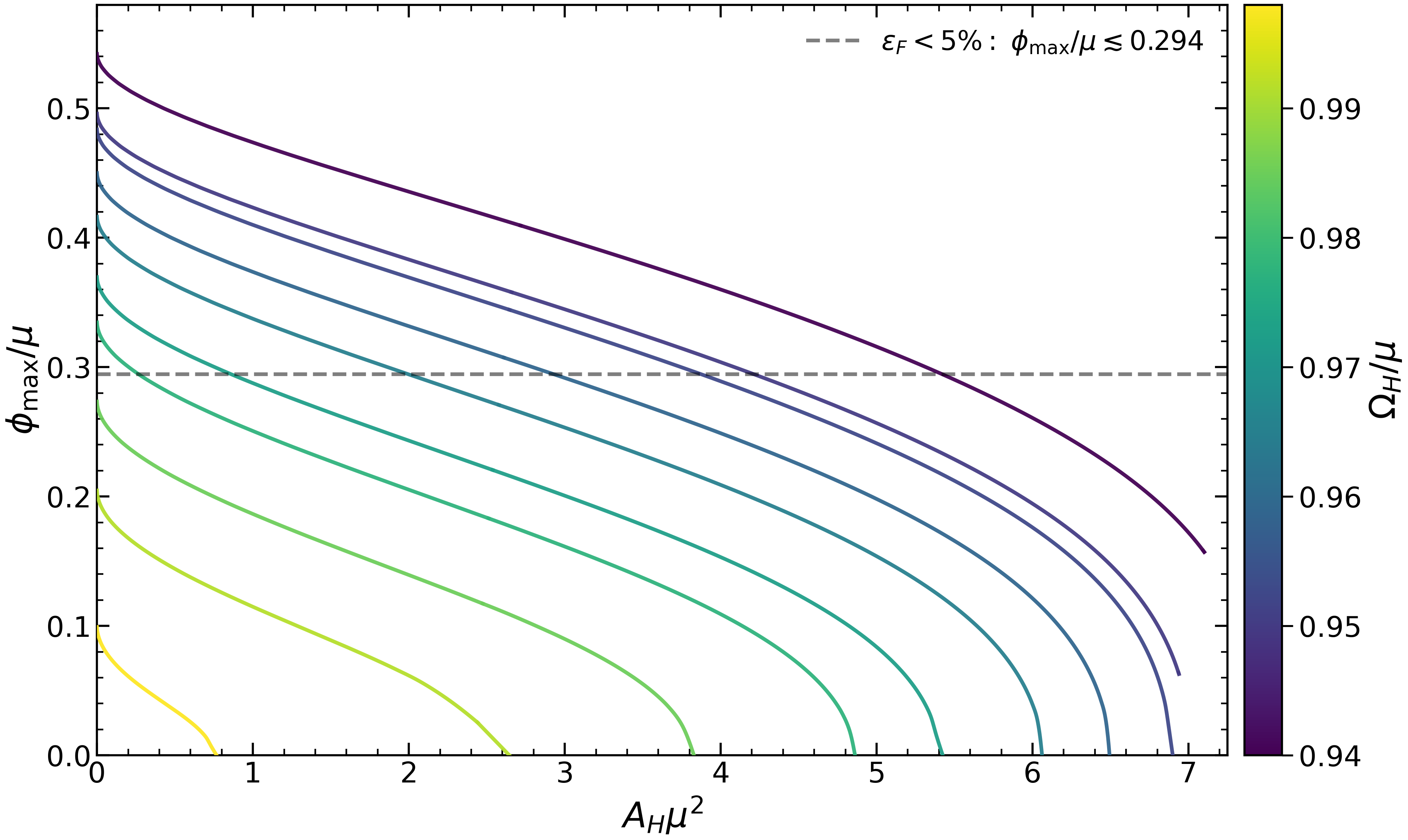}
    \caption{Maximum scalar amplitude
    \(\phi_{\rm max}/\mu\) as a function of the dimensionless horizon area
    \(A_H\mu^2\) for selected quartic-cloud branches with \(m=1\).
    The selected branches lie close to the upper frequency boundary,
    \(\Omega_H/\mu\simeq 1\). The dashed line marks the amplitude below which the quartic and cosine scalar forces differ locally by less than five percent, \(\phi_{\rm max}/\mu\simeq0.294\).}
    \label{fig:phi4_phimax_validity}
\end{figure}

\begin{figure}[H]
    \centering
    \includegraphics[width=\columnwidth]{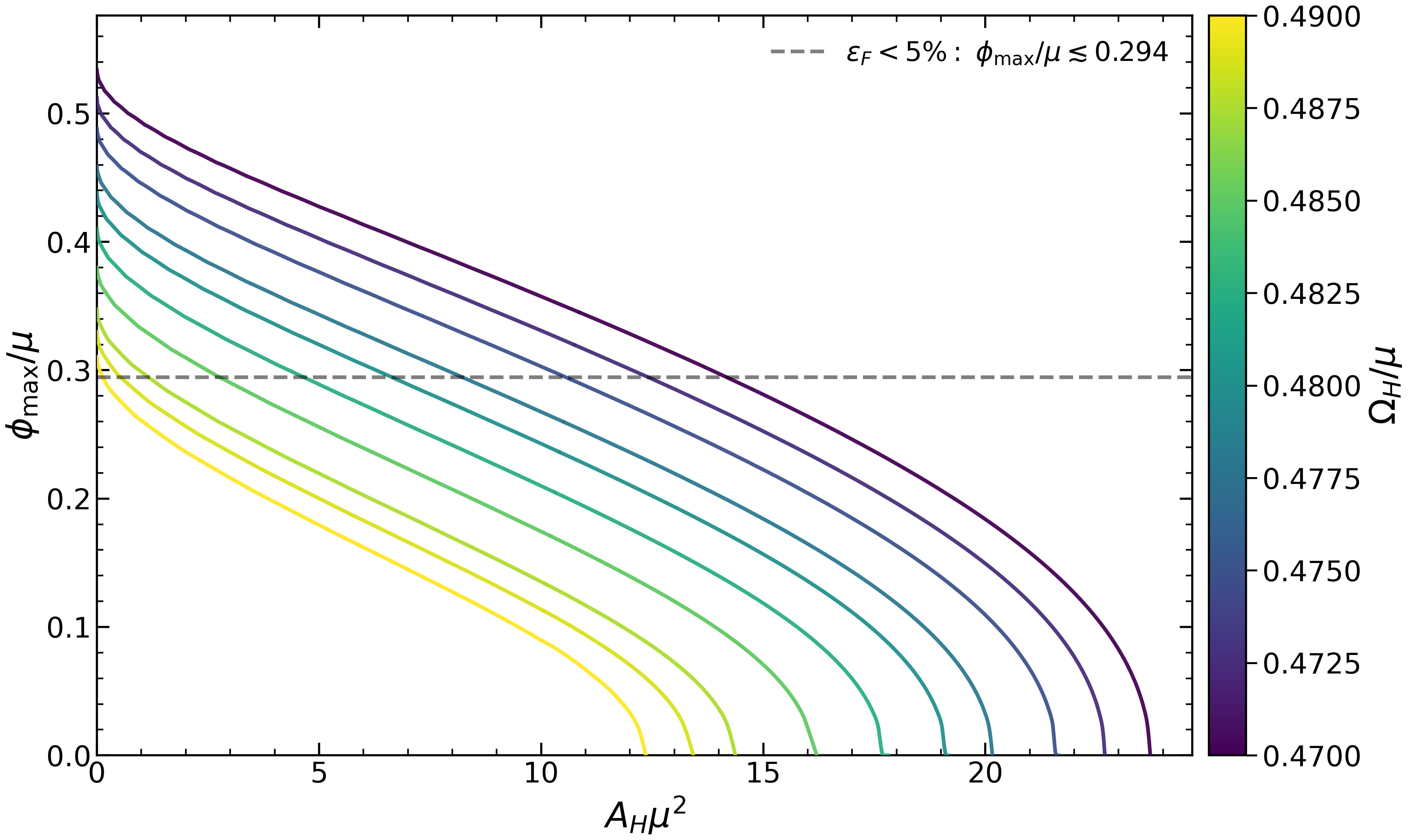}
    \caption{Same as Fig. \ref{fig:phi4_phimax_validity} for $m=2$.}
    \label{fig:phi4_phimax_validity_m2}
\end{figure}

Figs. \ref{fig:phi4_phimax_validity} and \ref{fig:phi4_phimax_validity_m2} display the maximum scalar amplitude along selected quartic-cloud branches with \(m=1,2\). The dashed horizontal line marks the threshold defined by Eq. (7), based on the relative deviation of the scalar force. Below this threshold, the quartic truncation provides a local approximation to the dynamics generated by the cosine potential. Above it, higher-order terms become quantitatively relevant, and the quartic solutions can no longer be interpreted as approximations to the axion-like model. They should instead be regarded as solutions of the quartic theory itself, whose unbounded large-field behaviour differs qualitatively from that of the bounded and periodic cosine potential. Consequently, features associated with the large-amplitude or low-frequency continuation of these branches need not persist in the full cosine model.

\begin{figure}[H]
    \centering
    \includegraphics[width=\columnwidth]{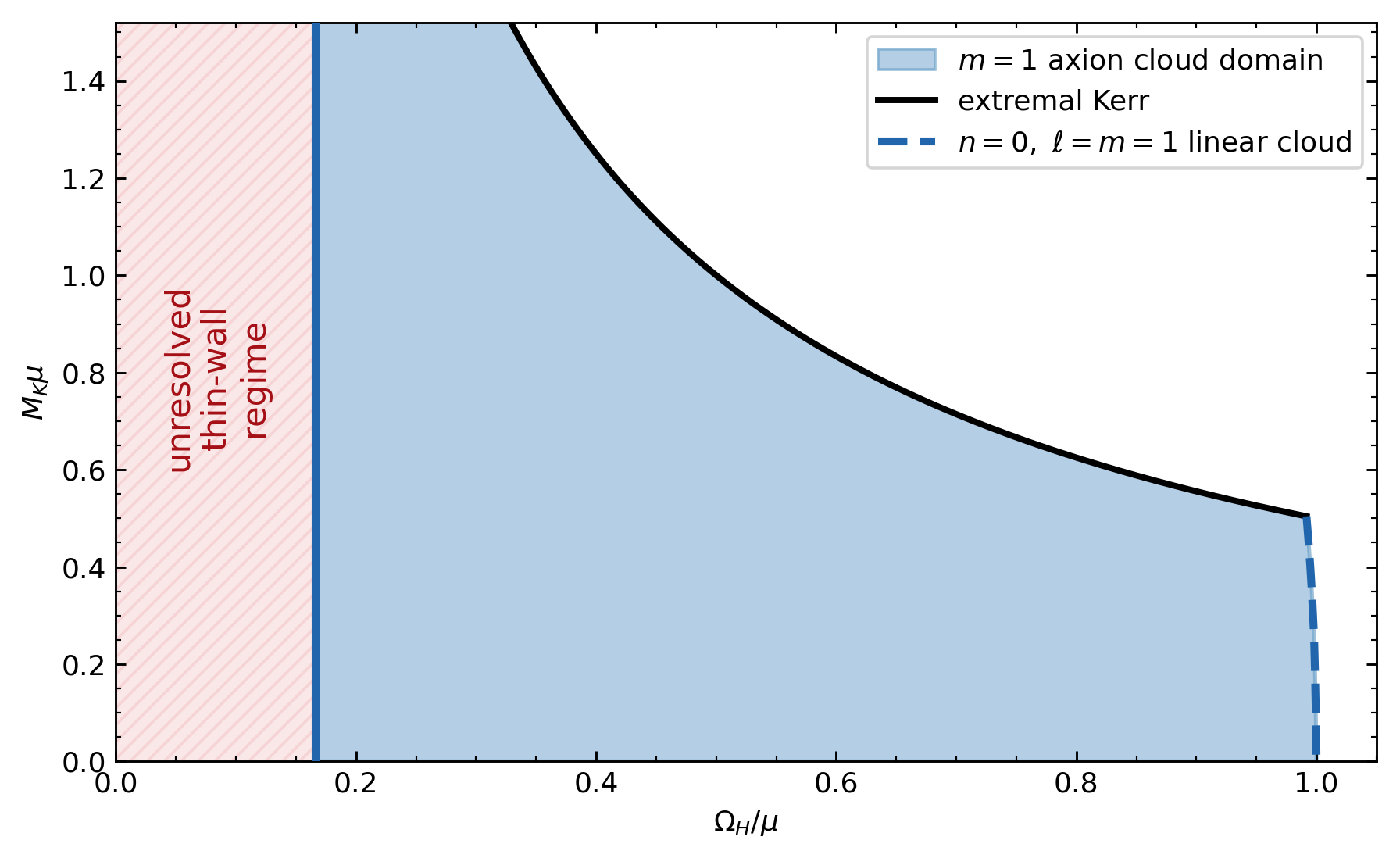}
    \caption{Existence domain for the normalised, axion-inspired cosine potential in the Kerr parameter space. The dashed curve denotes the linear cloud boundary, the black curve denotes extremal Kerr, and the vertical line marks the lowest resolved angular velocity, \(\Omega_H/\mu\simeq0.166\).}
    \label{fig:axion_domain}
\end{figure}

We also solved the fixed-background cloud problem for the full normalised
cosine potential in Eq.~\eqref{eq:cosine_completion}. This changes the
global domain structure. The numerical scan displayed in
Fig.~\ref{fig:axion_domain} shows  the
\((\Omega_H,M_K)\) plane. Its upper boundary is the linear cloud, where
the scalar amplitude tends to zero. Towards smaller angular velocity the solutions become increasingly broad and
approach a thin-wall-type regime. In this regime the scalar field remains close
to a nonzero value over an extended interior region and drops to the vacuum only
within a comparatively narrow transition layer. As \(\Omega_H\) is lowered, the
size of the cloud grows while this wall thickness remains much smaller than the
overall radius, producing a large separation of length scales and making the
continuation numerically stiff. In the scan shown here the lowest robustly
resolved value is \(\Omega_H/\mu\simeq0.166\). 

\begin{figure}[H]
    \centering
    \includegraphics[width=\columnwidth]{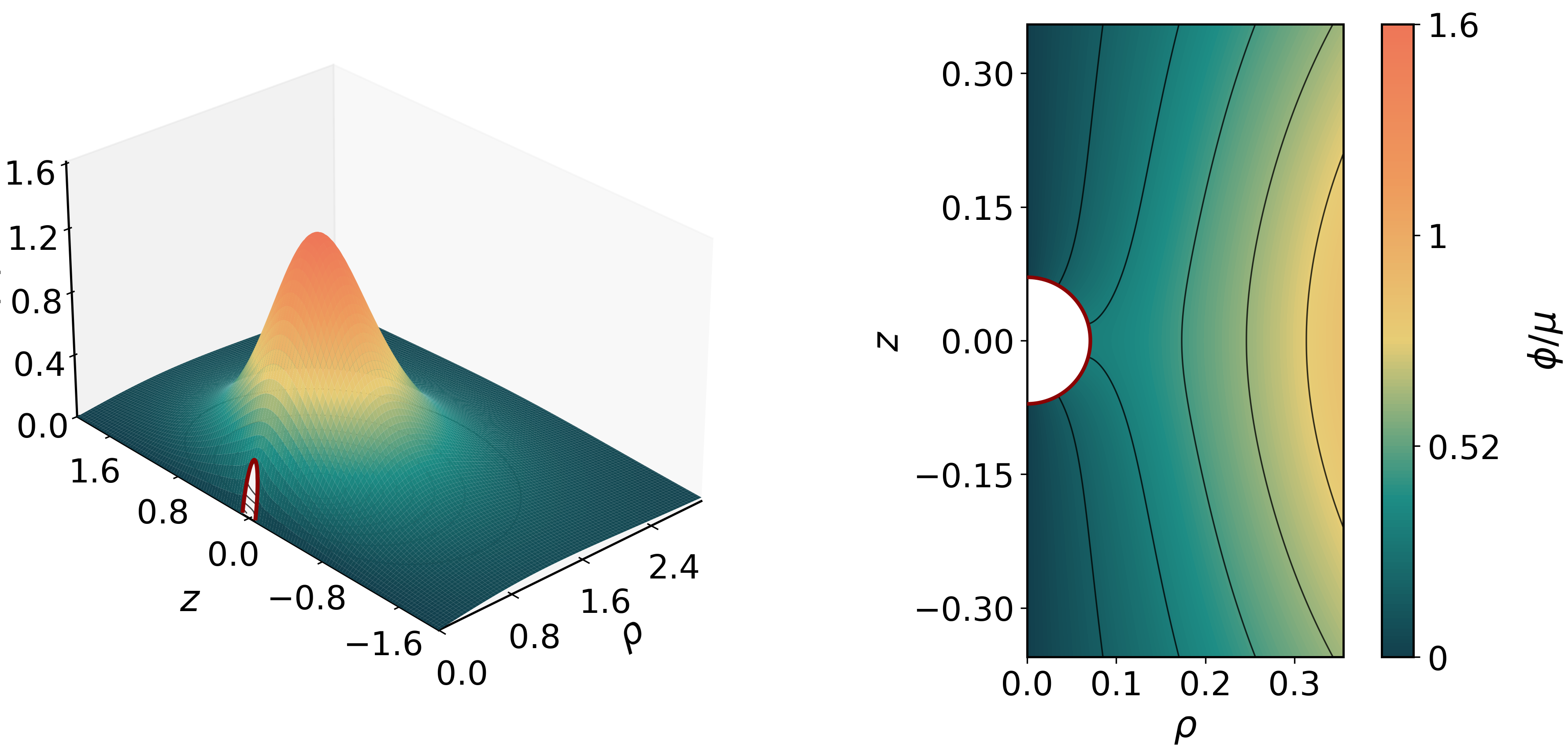}
    \caption{Scalar amplitude \(\phi\) of a representative quartic Q-cloud
    with \(m=1\), \(\Omega_H=0.5\), and \(A_H\mu^2=1.08\). The left panel shows
    the surface profile and the right panel a meridional heat map. The red
    semicircle marks the horizon.}
    \label{fig:kerr_phi_profile}
\end{figure}

\begin{figure}[H]
    \centering
    \includegraphics[width=\columnwidth]{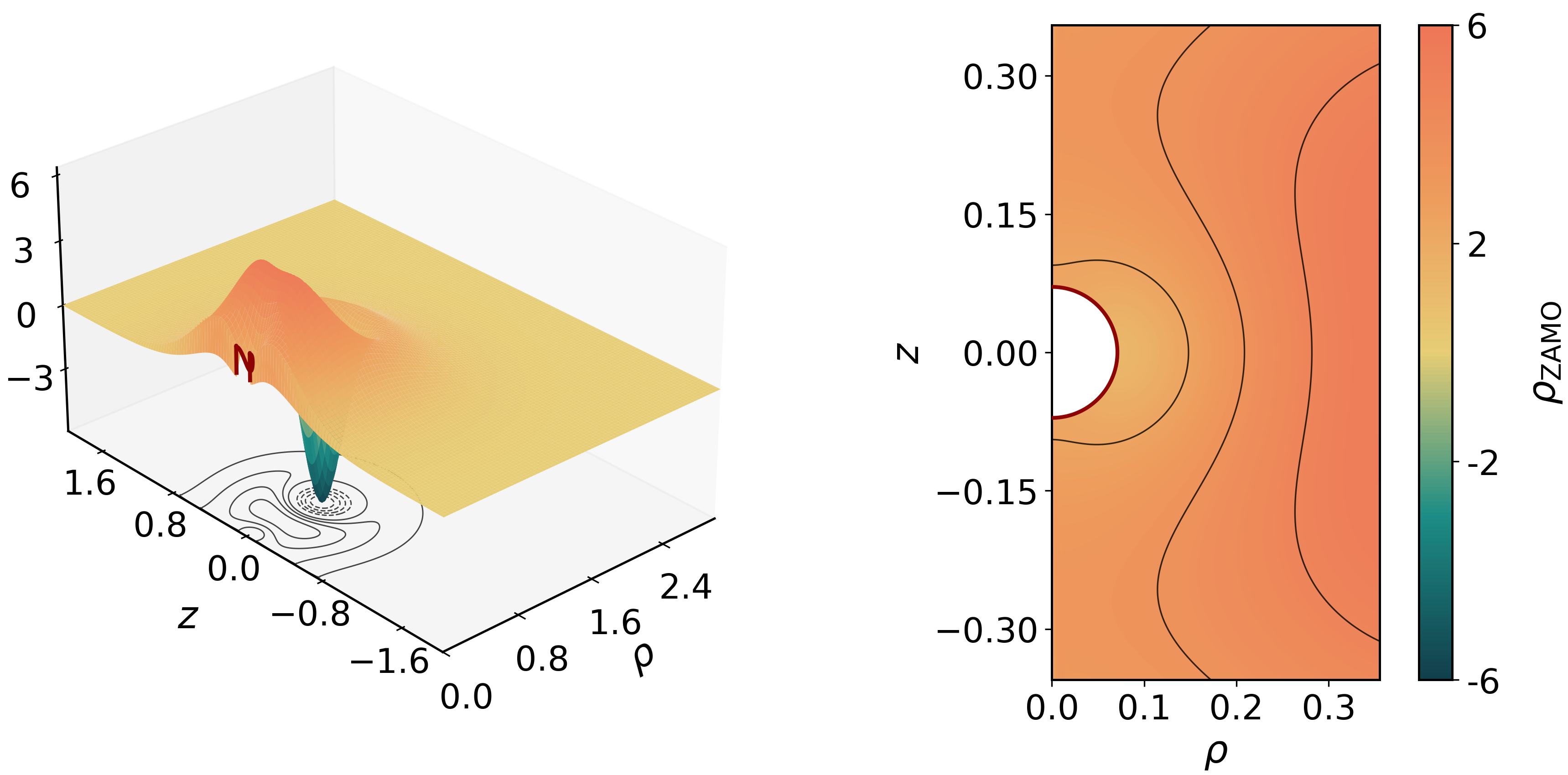}
    \caption{Same as Fig.~\ref{fig:kerr_phi_profile} for the ZAMO energy density \(\rho_{\rm ZAMO}\).}
    \label{fig:kerr_energy_density}
\end{figure}

\begin{figure}[H]
    \centering
    \includegraphics[width=\columnwidth]{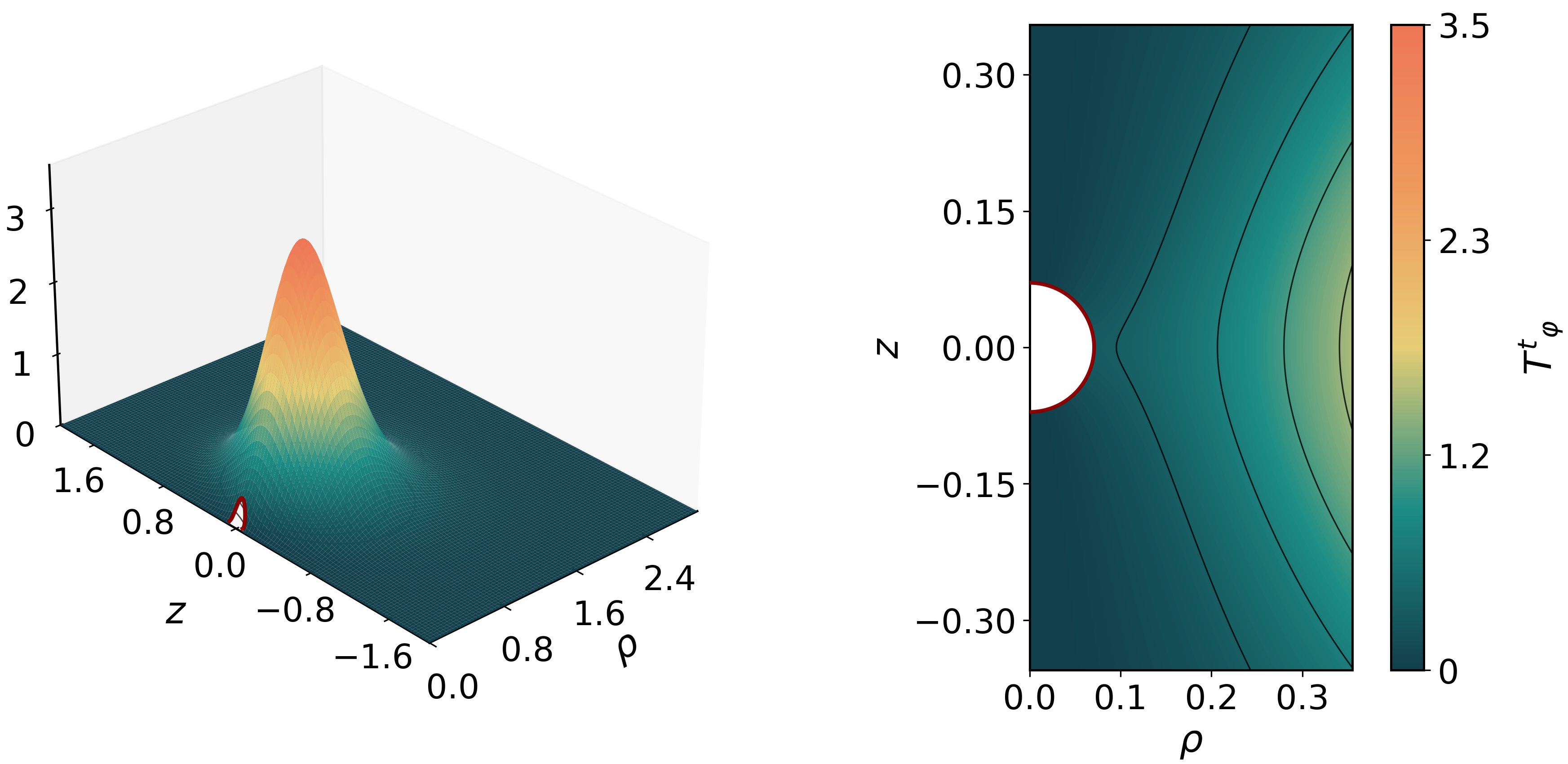}
    \caption{Same as Fig.~\ref{fig:kerr_phi_profile} for the angular momentum density \(T^t{}_{\varphi}\).}
    \label{fig:kerr_angular_momentum_density}
\end{figure}

Returning now to the quartic model, the structure of a representative \(m=1\) solution is displayed in
Figs.~\ref{fig:kerr_phi_profile}--\ref{fig:kerr_angular_momentum_density}.
We choose \(\Omega_H=0.5\) and the quasi-isotropic horizon area
\(A_H \mu^2=1.08\). As for spinning Q-balls and for the sextic Q-clouds of
Ref.~\cite{Herdeiro:2014pka}, the scalar amplitude is concentrated in a
toroidal region around the rotation axis. In a meridional section this
appears as a maximum at finite cylindrical radius \(\rho\). The horizon
removes the central part of the flat-space soliton, while regularity replaces
the condition imposed at the Minkowski origin by a horizon boundary
condition. The ZAMO energy density contains a negative region, as in the flat-space case. The angular momentum density remains localised in the same toroidal region.

\begin{figure}[H]
    \centering
    \includegraphics[width=\columnwidth]{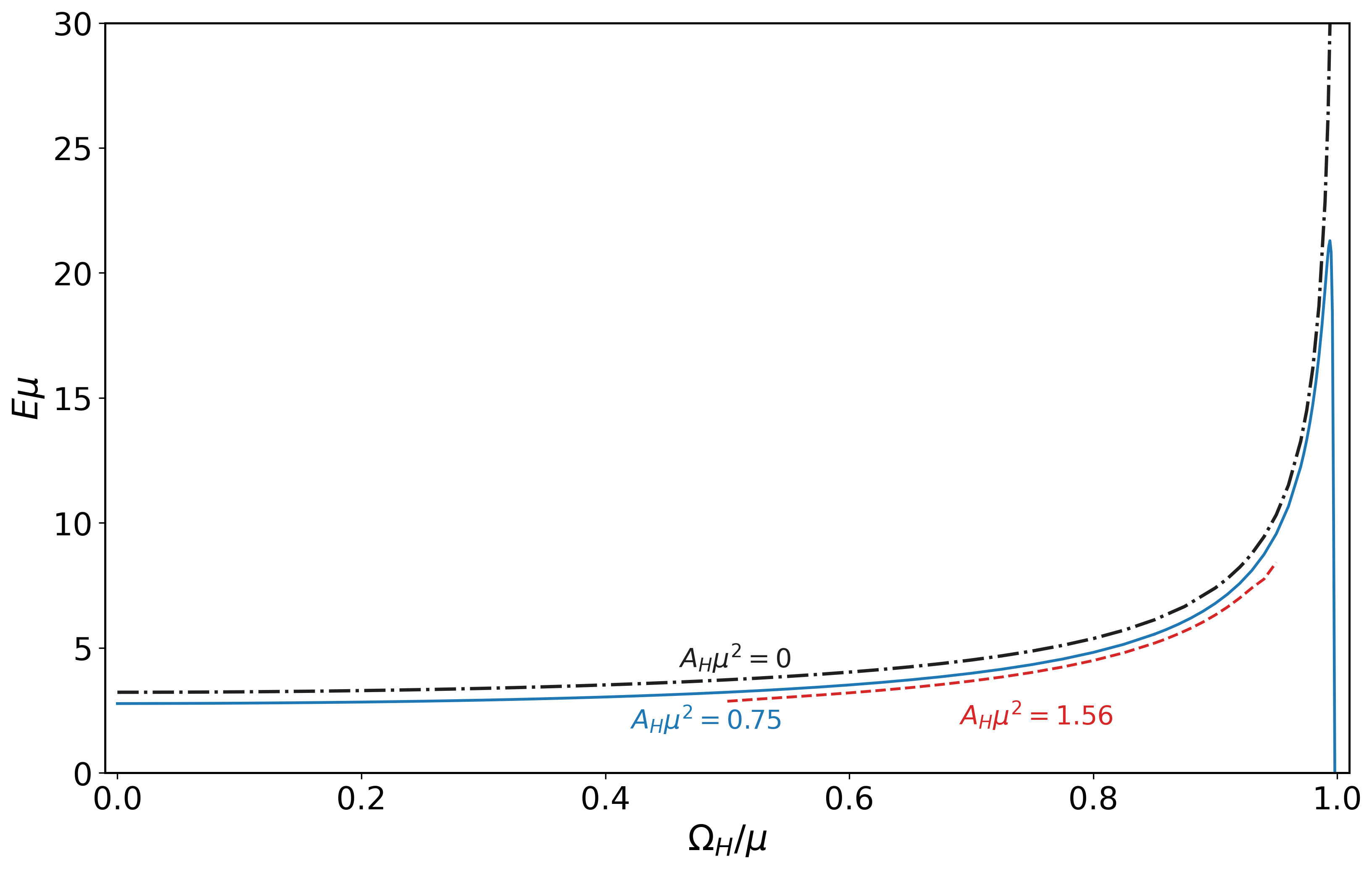}
    \caption{Energy \(E\) of the \(m=1\) quartic clouds as a function of
    \(\Omega_H\). The dash-dotted black curve is the flat-space limit
    \(A_H\mu^2=0\). The blue and dashed red curves correspond to fixed
    quasi-isotropic horizon areas.}
    \label{fig:kerr_branch_energy}
\end{figure}

Fig. \ref{fig:kerr_branch_energy} compares the energy spectrum of the
\(m=1\) quartic solutions in flat spacetime with two finite horizon Kerr
families. The black dash dotted line is the horizonless solitonic limit. The blue and
red curves are slices at fixed quasi isotropic horizon areas \(A\mu^2=0.75\) and
\(A\mu^2=1.56\), respectively. The cloud becomes increasingly extended as
\(\Omega_H\rightarrow\mu\), and the flat-space energy grows rapidly. Over
the  range of frequencies the finite horizon curves lie below the
flat space curve. In this sense, inserting a horizon excises part of the
soliton and lowers the integrated scalar energy. For some of the solutions found, the energy
first rises, reaches a narrow maximum, and then decreases sharply as the
solution approaches the fundamental linear cloud existence line. At this
boundary the scalar amplitude and the global scalar charges tend to zero.
The finite Kerr horizon therefore regulates the flat space divergence. This
is the same qualitative mechanism identified for the stabilised sextic
Q-clouds in Ref.~\cite{Herdeiro:2014pka}. 

\begin{figure}[H]
    \centering
    \includegraphics[width=\columnwidth]{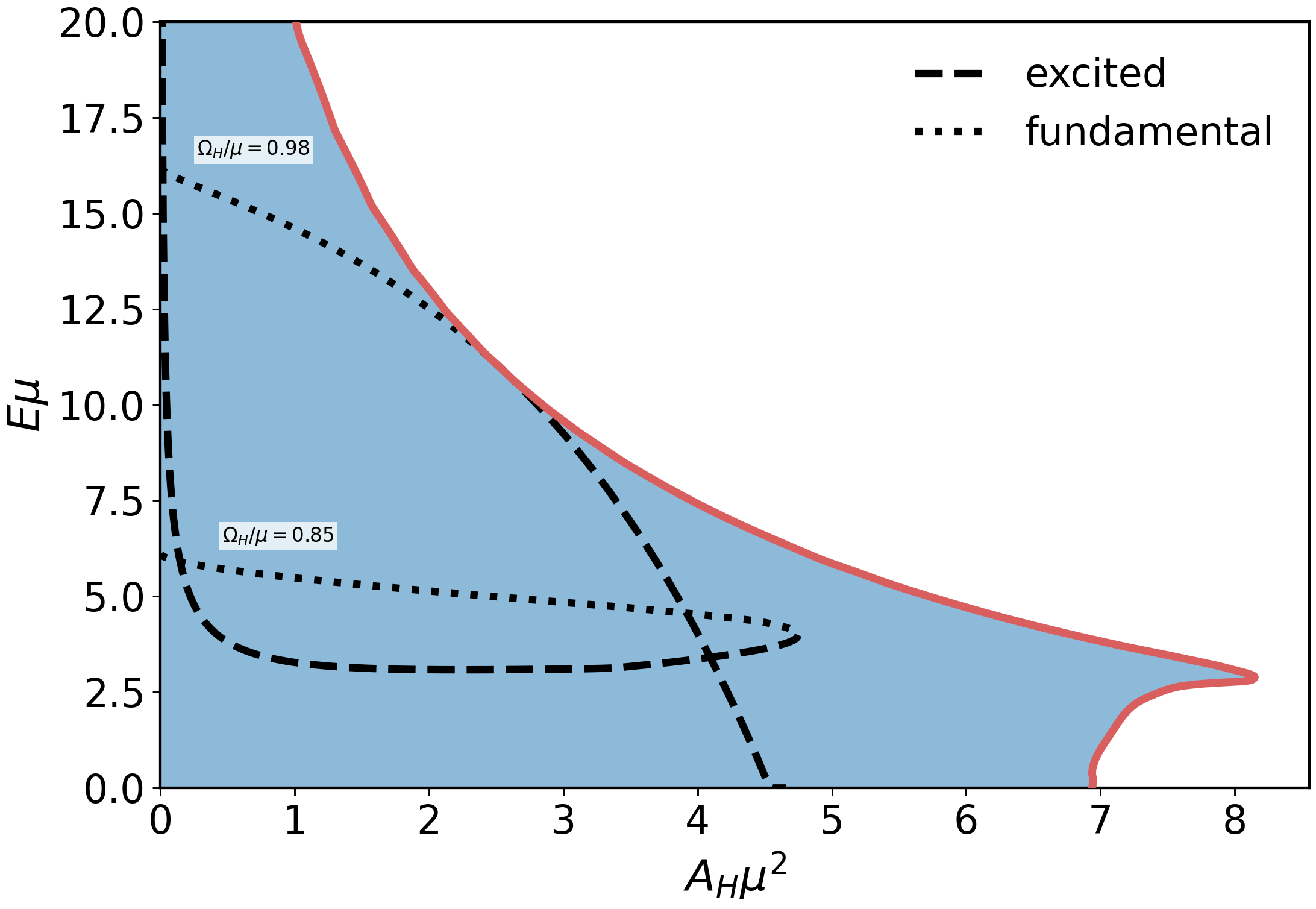}
   \caption{Dimensionless energy \(E\mu\) of
   selected \(m=1\) quartic-cloud branches as functions of the dimensionless horizon area
   \(A_H\mu^2\).The dotted and dashed black curves show the lower- (fundermental) and upper (excited)-mass geometric Kerr-branches, \(M_K^{\rm low}\) and \(M_K^{\rm up}\), respectively. The red
   curve is the numerically determined outer critical line of the domain.}
    \label{fig:parameter_space}
\end{figure}

\begin{figure}[H]
    \centering
    \includegraphics[width=\columnwidth]{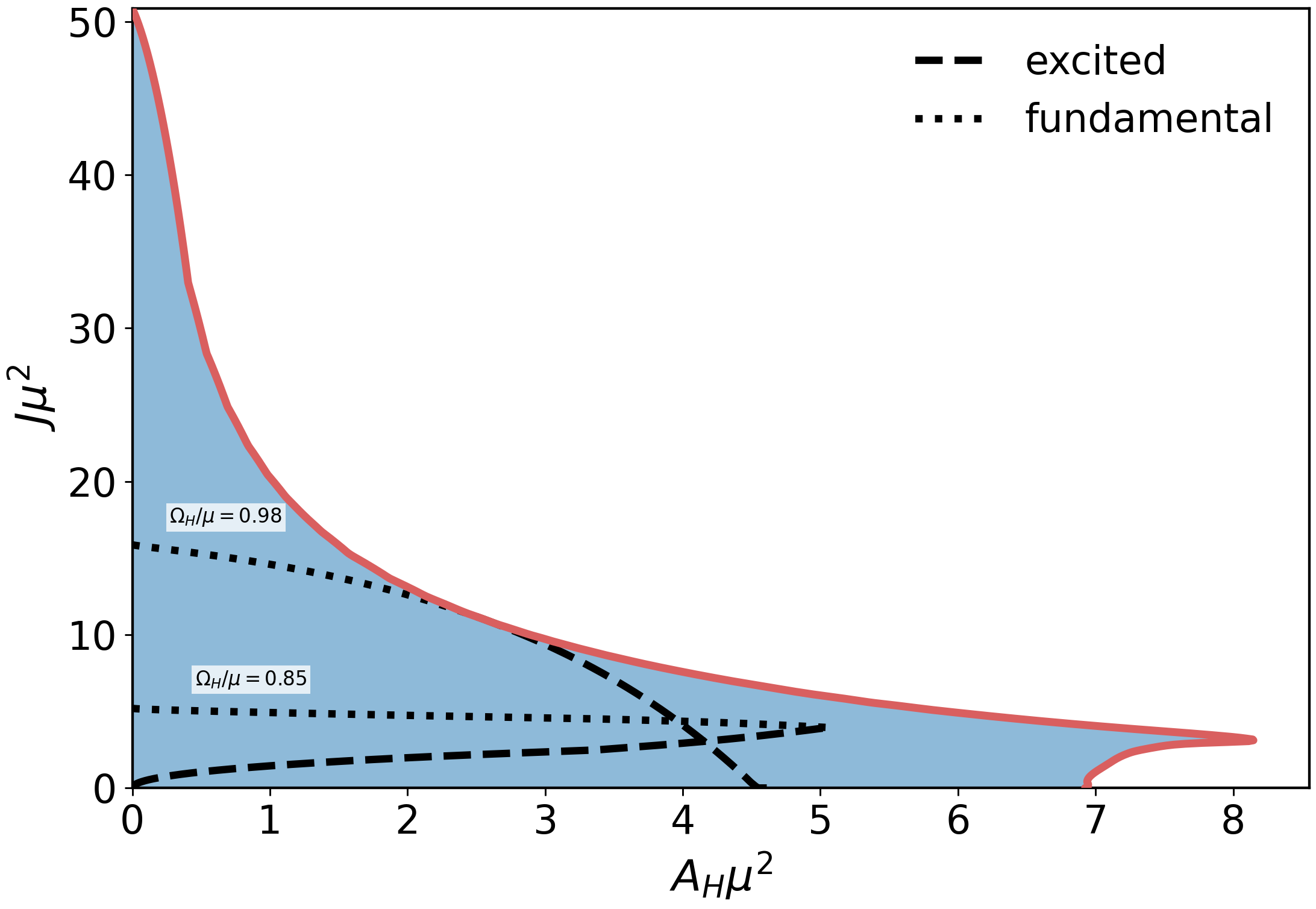}
   \caption{Same as Fig.~\ref{fig:parameter_space} for dimensionless angular momentum \(J\mu^2\).}
    \label{fig:parameter_space_angular_momentum}
\end{figure}

Finally let us show some quantities of interest obtained numerically. Figs. \ref{fig:parameter_space} and \ref{fig:parameter_space_angular_momentum} show the dimensionless energy \(E\mu\) and
angular momentum \(J\mu^2\) of selected \(m=1\) quartic Q-cloud solutions as
functions of the dimensionless horizon area \(A_H\mu^2\). For fixed
\(\Omega_H\), the solutions do not form a single-valued monotonic curve.
Starting from the first branch, the horizon area increases until a turning
point is reached. The continuation then folds back in \(A_H\), giving rise to
a second branch. Consequently, several physically distinct scalar
configurations, with different profiles and global charges, may exist for the
same values of \(A_H\) and \(\Omega_H\).

\section{Conclusion and outlook}

We have studied nonlinear scalar clouds of a complex scalar field with an
attractive quartic self-interaction on fixed Schwarzschild and Kerr
backgrounds. Motivated by Gubser's observation that scalar hair may arise when
the potential is allowed to become negative, the quartic model provides a
simple setting in which the usual positivity assumptions are relaxed. The
model should be regarded as an effective theory that isolates the leading
attractive interaction while omitting stabilising higher-order terms. In flat spacetime, the absence of a positive lower frequency bound allows the
quartic branches to extend to \(\omega=0\). On Schwarzschild backgrounds, this
gives static nonlinear clouds with \(Q=J=0\). On Kerr backgrounds, the
interaction opens the linear-cloud existence lines into two-dimensional
Q-cloud domains satisfying \(\omega=m\Omega_H\). In contrast to stabilised
sextic models, these domains have no minimum angular velocity imposed by the
potential, can connect to the \(\Omega_H=0\) Schwarzschild sector, and are not
globally bounded by the extremal Kerr curve. Solving the full cosine model directly produces a different global domain, with increasingly broad solutions and a numerically stiff thin-wall-type  Future work should include gravitational backreaction and an analysis of the dynamical stability of both classes of solutions.

\section*{Acknowledgments}

The author would like to thank Jutta Kunz and Eugen Radu for their valuable support and supervision throughout this work.

\newpage

\bibliographystyle{unsrt}   
\bibliography{biblio}

\end{document}